\documentclass[a4paper,12pt]{article}    %Type de document On peut ajouter twoside, twocolumn
\usepackage{a4wide}             % Pour avoir de petites marges
\usepackage{fancyhdr}           % Pour l'ent{\^e}te et le pied de page
\usepackage{amstext}
\usepackage{latexsym}           % Utilisation de la police LaTeX
\usepackage{amssymb,amsmath,amsthm}  % Pour les polices math{\'e}matiques et les th{\'e}or{\`e}mes
\usepackage{mathrsfs}                % Pour les polices math{\'e}matiques
\usepackage{amsfonts}
\usepackage{euscript}
\usepackage[ansinew]{inputenc}  % Sp{\'e}cifie le codage des caract{\`e}res utilis{\'e}s
\usepackage{textcomp}           % Il y a un Warning si on ne l'utilise pas
\usepackage[english]{babel}    % Langue du document
\usepackage{times}              % Un style assez joli
\usepackage{float}              %Pour mettre les figures ici [H]
\usepackage{lastpage}           %Permet de conna{\^\i}tre la derni{\`e}re page
\usepackage{multirow}
\usepackage{mhequ}
\usepackage{url}
\usepackage{wrapfig} % Enveloppement des figures \begin{wrapfigure}{align}{largeur} \end{wrapfigure}
\usepackage{subfig} % Pour les sous-figures
\usepackage{psfrag}%pour modifier le texte {\`a} l'int{\'e}rieur des fichiers eps
\usepackage{color}%Pour mettre des couleurs
\usepackage[dvips]{graphicx}
\usepackage{cite}

\DeclareGraphicsExtensions{.eps,.ps}

\date{\today}            %Indique la date manuellement

\title{Leader neurons in leaky integrate and fire neural network simulations}      %Indique le titre du document

\author{Cyrille \textsc{Zbinden}\\[3mm]
D{\'e}partement de physique th{\'e}orique\\
Universit{\'e} de Gen{\`e}ve\\
CH-1211 Gen{\`e}ve 4\\
Switzerland\\
E-mail: cyrille.zbinden@unige.ch}        %Indique l'auteur du document

\newcommand{\real}{{\rm I\kern-0.21em R }}
\newcommand{\integer}{{\rm I\kern-0.18em N }}

\definecolor{brun}{rgb}{0.8,0.4,0}

\makeatletter
\newlength{\earraycolsep}
\def\eqnarray{\stepcounter{equation}\let\@currentlabel%
\theequation \global\@eqnswtrue\m@th
\global\@eqcnt\z@\tabskip\@centering\let\\\@eqncr
$%
$\halign to\displaywidth\bgroup\@eqnsel\hskip\@centering
$\displaystyle\tabskip\z@{##}$&\global\@eqcnt\@ne \hskip
2\earraycolsep \hfil$\displaystyle{##}$\hfil &\global\@eqcnt\tw@
\hskip 2\earraycolsep $\displaystyle\tabskip\z@{##}$\hfil
\tabskip\@centering&\llap{##}\tabskip\z@\cr} \makeatother
\def\r{{\mathsf r}}
\def\w{{\mathsf w}}
\def\b{{\rm burst}}
\def\p{{\rm isolated}}
\def\BB{{\cal B}}
\def\PP{{\cal P}}
\def\QQ{{\cal Q}}
\def\RR{{\cal R}}

\def\ie{{\it i.e.}}

\newcommand{\captionfonts}{\small}

\makeatletter  % Allow the use of @ in command names
\long\def\@makecaption#1#2{%
  \vskip\abovecaptionskip
  \sbox\@tempboxa{{\captionfonts #1: #2}}%
  \ifdim \wd\@tempboxa >\hsize
    {\captionfonts #1: #2\par}
  \else
    \hbox to\hsize{\hfil\box\@tempboxa\hfil}%
  \fi
  \vskip\belowcaptionskip}
\makeatother   % Cancel the effect of \makeatletter

\begin{document}

\maketitle

\begin{abstract}\addcontentsline{toc}{section}{Abstract}

In this paper, we highlight the topological properties of leader neurons whose existence is an experimental fact.

Several experimental studies show the existence of leader neurons in
population bursts of 2D living neural networks
\cite{LeaderMarom,Leadership}. A leader neuron is, basically, a
neuron which fires at the beginning of a burst (respectively
network spike) more often that we expect by looking at its whole
mean neural activity. This means that leader neurons have some burst
triggering power beyond a simple statistical effect. In this study,
we characterize these leader neuron properties. This naturally
leads us to simulate neural 2D networks. To build our simulations,
we choose the leaky integrate and fire (lIF) neuron model \cite{GerstnerKistler,Cessac}, which
allows fast simulations \cite{IzhiWhich,WulfCompare}.

Our lIF model has got stable leader neurons in the burst population that we simulate. These leader neurons are excitatory neurons and have a low membrane potential firing threshold. Except for
these two first properties, the conditions required for a neuron
to be a leader neuron are difficult to identify and seem to depend
on several parameters involved in the simulations themself. However,
a detailed linear analysis shows a trend of the properties
required for a neuron to be a leader neuron. Our main finding is:
A leader neuron sends a signal to many excitatory neurons as
well as to a few inhibitory neurons and a leader neuron receives
only a few signals from other excitatory neurons.

Our linear analysis exhibits five essential properties for leader
neurons with relative importance. This means that considering a
given neural network with a fixed mean number of connections per neuron, our analysis gives us a way of predicting
which neuron can be a good leader neuron and which cannot. Our
prediction formula gives us a good statistical prediction even if,
considering a single given neuron, the success rate does not reach
hundred percent.

\end{abstract}

\newpage
\tableofcontents
\newpage

\section{Introduction}\label{Intro}

Disassociated \emph{in vitro} rat brain neuron cultures show, under a variety of
experimental contexts, a spontaneous electrical
activity. This activity manifests itself by a rapid succession of ignitions of a large
fraction of neurons named collective bursts (or network spikes)
\cite{TscherterInitiation, MaedaGeneration1995, droge1986mac,
PotterDevelopment}. This bursting activity can be measured, for example, by multi-electrode array methods \cite{LeaderMarom,Leadership,Wagenaar2006} or by fluorescence \cite{JordiDevelopment}. The study of the initiators of these bursts is one of the conceptual problems underlying the spontaneous
electrical activity.
A recent study \cite{LeaderMarom} showed that
some ``first to fire'' cells exist. Later \cite{Leadership}, through a detailed analysis of data obtained by the multi-electrode array methods, established that some cells are triggering bursts beyond a simple statistical effect of being the first. These particular cells are
called \emph{leaders}. The analysis \cite{Leadership} indicates that the long
term dynamics of the leaders is relatively robust, evolving with a half-life of
about one day. \cite{Leadership} also shows that these leaders are not only the main initiators of bursts, but, the burst itself carries traces (or hints) indicating which of the leaders has
initiated the burst. In this respect, one can view the culture as
an amplifier of the signal emitted by the leaders.

It is clear that the scenario described above calls for a theoretical
explanation. One of the ways pursued is the one of (bootstrap)
percolation in random networks \cite{Bootstrap} which gives insights into
the spread of the initial ignition and distinguishes between localized
and de-localized spreading. In this picture, the burst
itself is viewed as the ``giant component'' of the percolation
process, and this picture is verified from many different angles
\cite{JordiDevelopment}.

In this paper, we address the delicate question of what
makes a neuron a leader. Is it stimulation? Activity threshold?
Special connectivity properties of the network? The natural way of reaching this
aim theoretically is to simulate a random neural network. Our results show
that leadership is an effect of a combination of several ingredients
which can be quantified by a simple relation between several natural
parameters of the neuron. We obtain these results by a simple model
of a randomly connected network with the usual leaky integrate and fire \cite{GerstnerKistler} mechanism for ignition of neurons. This well known model \cite{Cessac} is tested in
numerical event-based simulations that provides, in principle, unbiased simulations \cite{CessacNumerical}. This model only takes into account simple properties of neurons (integrate and fire) \cite{IzhiWhich,WulfCompare} but it suffices to reproduce basically all the findings of \cite{Leadership}. Thus we trade realism for conceptual simplicity, but in this kind of investigations the complexity comes from the neural network itself and not from the neuron model.

The results of our simulations can be presented as follows: to each neuron one can assign a
leadership score and neurons with the highest scores are leaders. It turns out that leaders can be characterized as being excitatory neurons and having a low membrane potential firing threshold. Apart from these two first properties, the conditions required for a neuron to be a leader are difficult to identify and depend
on several parameters involved in the simulations themself. Indeed,
a detailed linear analysis shows a trend of the properties
required for a neuron to be a leader. The main finding of this paper is 
a formula for the leadership score that exhibits five essential properties for leaders with relative importance. This formula gives a very good statistical prediction. Therefore, we can conclude that leadership is a random effect, and that leaders are formed naturally from a balanced combination of inputs, outputs,
their local neighborhood and their own properties. Basically a leader sends a signal to a lot of excitatory neurons as well as to a few inhibitory neurons and a leader receives
only a few signals from other excitatory neurons.

\section{The simulations}\label{simulation}

In this section we explain how we construct our leaky integrate and fire
simulations and which kind of parameters we use.

\subsection{Building the simulations}\label{strategy}

A single neuron is already a complex biological object. But, basically, neurons are electrically excitable cells that are composed of a soma (cell body), a dendritic tree and an axon. Dendrites and axons connect to each other (through synapses) to create a neural network. Thus, to make a model, we can, for example, choose a geometry and some rules that produce the rule of connection between neurons \cite{MaThese} or construct a random
neural network with a fix mean number of connections per neuron.

\begin{figure}[ht]
\begin{center}
\parbox{0.5\textwidth}{\includegraphics[width=0.45\textwidth]{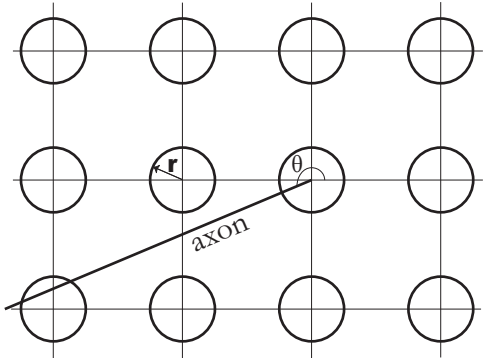}}\parbox{0.5\textwidth}%
{\caption[Geometry of the network]{A $3\times 4$ neurons
rectangle. Each neuron is a circle with a radius $\r\in\real_+$
(dendrites), in this figure $\r<1$. The axon of one of the neurons
is drawn. We see that this neuron is connected to the neuron on
the left of the bottom row. Note that the axon direction is given
by $\theta\in[0,\ 2\pi)$. Note also that if $\r\geq1$, unlike in this
figure, then we automatically have nearest neighbour
connections.}\label{FigGeometry}}
\end{center}
\end{figure}

Our simulations are built as follows. First we construct a matrix of synaptic weights $W$ that represents a random neural network of $N\in\integer$ neurons where $N$ is a parameter and $\sqrt{N}\in\integer$ as so we can consider that the neurons are placed on a grid $\left\{(x,y)\in\integer^2|\ x,y\leq
\sqrt{N}\right\}\subset\real^2$ with periodic boundary conditions. In the Euclidean geometry of $\real^2$,
we consider that every neuron $n$ $\left(n=1,2,\ldots,N\right)$ is a circle of radius $\r\in\real_+$ (equivalent to the spatial extension of the dendrites). Each axon has a length $\ell\in\real_+$, approximately Gaussian distributed (with probability density function close\footnote{Negative axon lengths are not allowed and  we impose a maximum length of $\frac{2}{3}\sqrt{N}$ for each axon. With this last condition, no
axon can cover all the network in any direction. The rationale for
this choice is that, in the experiments \cite{Leadership}, probably no axon covers
all the test tube in a given direction.} to $f(\ell,\ L)=\frac{1}{\lambda\sqrt{2\pi}} e^{-\frac{(\ell-L)^2}{2\lambda^2}}$ where $L\in\real_+$ is the mean axon length and $\lambda=\frac{1}{3}\textrm{min}\left(L-1,\frac{2}{3}\sqrt{N}-L\right)$ is the standard deviation (and $1<L<\frac{2}{3}\sqrt{N}$)). In our simulations, the neuron spatial parameters are the dendrite size
$\r$ and the mean axon length $L$. To every axon, we also
associate a direction (respectively an angle) $\theta\in[0,\
2\pi)$ uniformly distributed (with probability density function
$g(\theta)=\frac{1}{2\pi}$ ). Figure \ref{FigGeometry} illustrates the network geometry.

To construct the matrix $W\in\textrm{M}_{N}(\real)$, we proceed as follows: If the axon
(with finite length $\ell$, without width and angle $\theta$
(respectively direction)) of the neuron $m$ $(m=1,\ldots,N)$
intersects the circle (dendrites) of the neuron $n\neq m$ $(n=1,\ldots,N)$
then the connection exists (\ie $W_{nm}\neq0$), otherwise
$W_{nm}=0$. The biological knowledge tells us that the connection
between neurons is oriented (electrical impulses move from axons
to dendrites). Thus we have a non-symmetric matrix of synaptic
weights $W$ representing the neural network.

A fix parameter $\w\in\real_+$ named the synaptic weight is the value we give to $W_{nm}$ when the connection exists (\ie $W_{nm}\neq0$). Note that, in reality, some neurons are inhibitors. For
these, we use a negative synaptic weight $-\w\in\real_-$. In our simulations the proportion $r\in[0,\ 1]$
of inhibitory neurons is also a parameter.

To simplify the terminology, if $W_{nm}\neq0$, we will say that the neuron $n$ is
the son of the neuron $m$ and the neuron $m$ is the father of the
neuron $n$. One given neuron can have more than one father and
more than one son. Of course, a given neuron can be both, father and son of another neuron.

Once the matrix of synaptic weights is constructed we assign two real values to each neuron $n$: $V^{*}_{n}$ the membrane potential firing threshold and $V_n(t)$ the membrane potential which depends on the time $t\in\real_+$. Without loss of generality, we can consider that the membrane potential
\footnote{We know that the membrane potential oscillates more or less between $-70$ and $-50$ mV \cite{GerstnerKistler}. But to simplify, we will approximately scale the membrane potential between $0$ and $1$.} oscillates roughly between $0$ and $\langle V^{*}\rangle=1$ (where $\langle V^{*}\rangle$ is the mean of all the neuron membrane potential firing thresholds). The time evolution of the membrane potential $V_n(t)$ is computed using the leaky integrate and fire \cite{GerstnerKistler} neuron model.

The dynamics is the following, every neuron $n$ (respectively every membrane potential $V_n(t)$) which integrates the electrical signal it receives from the other neurons is able to fire (when $V_n(t)>V^{*}_{n}$) and then sends a signal to the other neurons. Just after firing the membrane potential is reset to $0$. The membrane potential also decreases with a leaky time $\tau$ (loss of memory).

More precisely, the membrane potential evolution equation of the neuron $n$ is the evolution equation of a leaky
integrate and fire neuron model \cite{GerstnerKistler}
\begin{equation}\label{VevolLIF}
V_n(t+ dt)=\underbrace{V_n(t)e^{-\frac{dt}{\tau}}}_{\textrm{Loss of memory}}\underbrace{\left(1-\chi_n\left[V_n(t)\right]\right)}_{\textrm{Reset after firing}}
+\underbrace{\sum_{m=1}^{N}W_{nm}\chi_m\left[V_m\left(t-\frac{d_{nm}}{v}\right)\right]}_{\textrm{Connections}}
+\underbrace{B_{t,n}\cdot\Omega_{t,n}}_{\textrm{Noise}}
\end{equation}
where
\begin{itemize}
  \item $dt$ is the infinitesimal time step,
  \item $\tau\in\real_+$ is the leaky time (or characteristic time of the membrane \cite{GerstnerKistler}),
  \item $\chi_n[V_n(t)]$ is, for the neuron $n$, the indicator function of the set of
  potentials bigger or equal to $V_{n}^{*}$. Namely, $\chi_n[V_n(t)]=1$ whenever $V_n(t)\geq V_{n}^{*}$ and $\chi_n[V_n(t)]=0$ otherwise. The distribution of all the characteristic membrane potential firing threshold is approximately Gaussian with a probability density function close to
$h\left(V^{*}\right)=\frac{1}{\Delta V^{*}\sqrt{2\pi}}e^{-\frac{(V^{*}-1)^2}{2\Delta V^{*\ 2}}}$
where $\Delta V^{*}$, the standard deviation, is a parameter,
  \item $d_{nm}=d_{mn}$ is the distance between the neuron $n$ and the neuron $m$. 
In the experiments related in \cite{Leadership}, one finds a 2D density of
about $2000$ $\frac{\textrm{cells}}{\textrm{mm}^2}$. This leads us
here to fix the metric scale in our simulations: If we observe an expected mean number of neurons of $2000$ neurons inside the grid square, we assume that this square measures $1$ mm$^2$.
Nevertheless, we can also measure the distance in an arbitrary
unit given by the expected number of neurons in a row but we need the metric scale to compare our signal propagation velocity to the experimental one,
  \item $v$ is the propagation velocity of the signal in the neural network,
  \item $B_{t,n}$ is, for each time $t$ and each neuron $n$, a Gaussian random variable with mean
  $\sigma>0$ and standard deviation $\frac{\sigma}{3}$. Roughly speaking $B_{t,n}\sim \mathcal{N}\left(\sigma,\frac{\sigma}{3}\right)$ for all $n$ and for all $t>0$. To generate this Gaussian noise, we use a Box-Muller transform,
  \item $\Omega_{t,n}$ is a Markovian exponential clock with a mean rate $\omega$.
  For most $t$, one has $\Omega_{t,n}=0$, but sometimes $\Omega_{t,n}=1$. More precisely, the distribution of the time intervals $\Delta t$ of the continuous set $t$ where $\Omega_{t,n}=0$ follows the exponential distribution $\frac{1}{\omega}e^{-\frac{\Delta t}{\omega}}$ ($\omega$ is the mean value). Finally, it means that $B_{t,n}\cdot\Omega_{t,n}$ is a noise $B_{t,n}$ which acts on the membrane potential of neuron $n$ only when the exponential clock $\Omega_{t,n}$ rings.
\end{itemize}

By looking at \eqref{VevolLIF}, we see that, in
principle, we can compute every membrane potential all the time.
But it is not useful and too complicated to compute every membrane
potential continuously. Choosing a fixed time step $\delta t>0$
also introduces some problems like synchronized events and
systematic errors \cite{CessacSam}. In the program, made in Perl,
we compute the membrane potential only when an event
happens\footnote{Here, by event we mean the moment when a neuron receives a signal from another neuron or when a neuron receives some noise (\ie the Markovian exponential clock
$\Omega_{t,n}$ rings).}. At this time, we check if the neuron
fires or not. So, in the program, we store the events to find out
the next membrane potential to update. In other words, our
simulations \eqref{VevolLIF} are event-based simulations. The
firing time of neurons is not discretized but computed event per
event at the machine precision level. This way of doing provides,
in principle, unbiased simulations \cite{CessacNumerical}.

Note that if ``every'' neuron fires at (almost) the same time
(this may happen in a burst) then all the membrane potentials are
reset to $0$ at the same time; the neurons are synchronized. To
avoid that this situation occurs too often, when we initialize a
sequence, we give to each neuron $n$ a membrane potential in $[0,\
V_n^{*})$ (uniformly distributed). Note that we reinitiate a
sequence each time a burst is too long, so it is possible that
during a simulation the neurons get synchronized.

Note also that if the mean exponential clock rate $\omega$ is too
small relatively to the leaky time $\tau$, it does not really matter
which kind of distribution we use for the noise $B_{t,n}$. This
means that if $\omega\ll\tau$ then we can use $B_{t,n}=\sigma$ for all $t$ and for all $n$ as the noise
distribution $B_{t,n}$ and the result of the simulations remains
the same.

\subsection{Values of parameters and simulation choices}\label{parameters}

The parameters are fixed in a way to observe activity like the one observed in the experiments \cite{Leadership} when we restrict the observation to sixty neurons. This section provide a
list of the order of magnitude of the parameters used in
\eqref{VevolLIF}. Note that the reader can find almost all the
information in \cite{RevueNeurones}.
\begin{itemize}
    \item We made simulations for different sizes $N\in\integer$. But beyond $N\sim4$ it does not affect the presence of bursts and leaders. This means that we already get a leader when we consider only four neurons.
    \item In fact fixing the dentrites size $\r$, the mean axon length $L$ and the axons length distribution fix the distribution function for the sons (out-degree) and the mean number of connections per neuron (noted $\langle\textrm{connections}/\textrm{neuron}\rangle$). Remark that the number of sons per neuron distribution looks like the axons length distribution while the number of fathers per neuron distribution is always approximately Gaussian distributed (data not shown). Note also that the geometry we choosed for our simulations induces that the successful pre-burst shows locality, like the one observed by the authors in \cite{Leadership}. Finally in most of our simulations, we chose $\r=0.85$ and $L=7$, then $\langle\textrm{connections}/\textrm{neuron}\rangle\cong11$.
    \item The synaptic weight $\w$ is surely a function of the neuron and of the time (the neuron's learning).  As soon as we choose mean number of connections per neuron and considering the kind of neural activity we want, we can fix $\w$. We use $\w\sim\frac{1}{10}$. Note that if $\w\cong\langle V^{*}\rangle=1$ (where $\langle V^{*}\rangle$ is the mean of all the neuron membrane potential firing thresholds) then all neurons are too strongly correlated and if $\w=0$ then all neurons are independent.
    \item For the inhibitory proportion $r$, the choice of $r=0.2$ is reasonable. Depending on the neural culture type, the inhibitory cells proportion can change approximately from $20\%$ to $30\%$ (see \cite{JordiDevelopment}).
    \item In our simulations, the leaky time $\tau$ is about $100$ ms. Note that the order of magnitude is the correct one \cite{MosesAlvarez}. To be exact we compute with $\tau=0.1$ s, this means that the used time unit is the second.
    \item The propagation speed $v$ in a neural network is between $50$ and $100$    $\frac{\textrm{mm}}{\textrm{s}}$, see \cite{RevueNeurones}. In our case, $v$ should be replaced by the propagation speed in the axon which is bigger. However, we have decided to use this range throughout.
    \item The standard deviation of the membrane potential firing threshold $\Delta V^{*}$ is also a parameter. We used $0\leq\Delta V^{*}\leq0.2\langle V^{*}\rangle$, which guarantees that neurons are quit similar and still the membrane potentials do not get synchronized too often.
    \item For the exponential clock rate we use $\omega$ between $0.001$ and $0.01$ s.
    \item The mean noise $\sigma$ is determined by the kind of neural activity we want to have (observe). If there is too much noise then we loose the locality in the successful pre-burst because we have a lot of uncorrelated activities. Note that the mean membrane potential (without connection) $\overline{V}=\lim_{T\rightarrow\infty}\frac{1}{T}\int_{0}^{T}V\left(\tilde{t}\right)d\tilde{t}$ is about\footnote{The mean membrane potential charge due to the noise is $\sigma+\sigma e^{-\frac{\omega}{\tau}}+\sigma e^{-2\frac{\omega}{\tau}}+\ldots=\frac{\sigma}{1-e^{-\frac{\omega}{\tau}}}\cong\frac{\tau\sigma}{\omega}$ if $\frac{\omega}{\tau}$ is small.} $\frac{\tau\sigma}{\omega}$. We want the mean membrane potential $\overline{V}$ to be high enough to get some neural activity due to the noise. But we also want that $\overline{V}+\w<\langle V^{*}\rangle$, so that one spike does not necessarily create a burst.

    One might think that to keep the neural activity one can vary the values $\omega$ and $\sigma$ keeping the ratio $\frac{\sigma}{\omega}$ constant because the mean membrane potential $\overline{V}$ is about $\frac{\tau\sigma}{\omega}$. But to preserve the neural activity, the mean membrane potential $\overline{V}$ is not that relevant. However it is necessary to keep constant the mean time the membrane potential of neurons need to reach the value $V^{*}$. From this point of view the relation between $\sigma$, $\omega$ and $\w$ is a first passage time problem. Note also that if the value of the mean membrane potential $\overline{V}$ is too low then there is no neural activity due to the noise (this means no neural activity at all). To conclude, even if the exact value of the mean membrane potential $\overline{V}$ is not fundamental, this value must not be too far from the mean of all neurons membrane potential firing threshold $\langle V^{*}\rangle$ (see \cite{GerstnerKistler}).
\end{itemize}
The parameters $N$, $\r$, $L$, $\Delta V^{*}$ and $r$ are the neural network parameters and the parameters $\tau$, $v$, $\w$, $\sigma$ and $\omega$ are the dynamical parameters.

\begin{table}[htb] %[hbt]
\begin{center}
\begin{tabular}{{l}|{c}|{c}|}
 & Neural network parameters & Dynamical parameters  \\
\hline \multirow{2}{*}{Known} & $r\cong20\%$ \small(inhibitory proportion)\small & $\tau\simeq100$ \small ms (leaky time)\small\\
& $\r$, $L$, \small(mean number of connections)\small & $v\cong100$ \small $\frac{\textrm{mm}}{\textrm{s}}$ (propagation speed)\small\\
\hline
\multirow{2}{*}{To be set} & $N$ \small(computing-memory)\small & $\langle V^{*}\rangle-\overline{V}>\w>0$ \small(synaptic weight)\small \\
& $\Delta V^{*}$ \small(firing threshold standard deviation)\small & $\sigma$, $\omega$ \small(noise parameters)\\
\hline
\end{tabular}
\end{center}
\caption[Known and unknown parameters of the simulations]{Known and unknown parameters of the simulations.
The unknown parameters will be set in Section
\ref{AboutParam}.} \label{tab:Param}
\end{table}

\section[General appearance of the simulations]{General appearance of the simulations and some definitions}\label{UsefullDef}

In this section we show what is the product of our simulations and we give
the definitions of what we call leader neuron, according to \cite{Leadership}. 

\subsection{General appearance}\label{generalappearance}

Like in the experiments \cite{LeaderMarom,Leadership,PotterDevelopment}, we want to look at the neural activity. The product of our simulations consists of a list of ordered spikes times. For each spike $i$, the list gives us $t_i$, the time of the $i^{\textrm{th}}$ spike, and $n_i$, the neuron that fired. Then, considering only the spike chronology, we analyze the ``spike sequence'' of a subset (or all) neurons of our simulations.
\begin{figure}[ht]
 \centering
    {\begin{psfrags}%
      \psfrag{15 [ms]:}[c][t][1]{\footnotesize{15 [ms]:}}%
    \includegraphics[width=0.9\textwidth, angle=0]{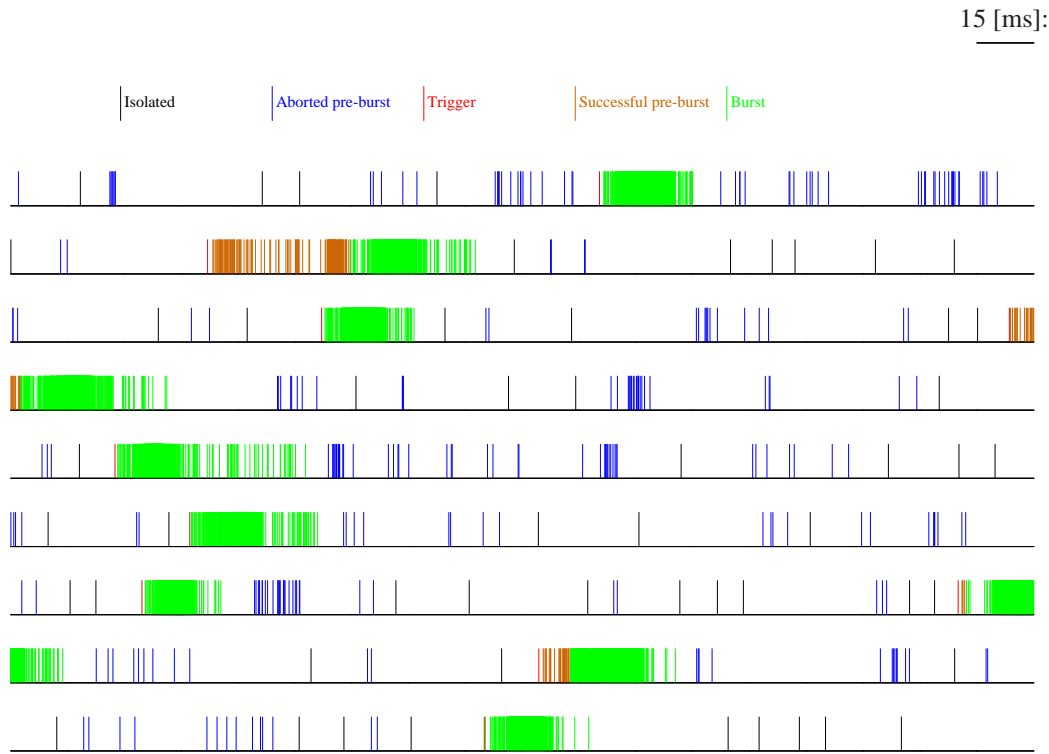}\end{psfrags}}\\
 \caption[Temporal evolution of a simulation with definitions]{Picture of the temporal evolution of a simulation for the whole neural activity with the parameters:
$N=900$, $\r=0.85$, $L=7$, $r=0.2$, $\Delta V^{*}=0.05$, $v=5$
$\frac{1}{\textrm{ms}}$, $\w=0.1$, $\tau=100$ ms, $\sigma=0.008$, $\omega=0.001$ s, $n_\b=500$, $\delta t_\b=15$ ms and $\delta t_{\p}=5$ ms.
The color code, that gives a representation of the different definitions explained in the following section, is the following: {\color{black} {\bf\emph{Isolated spike}}}, {\color{blue} \emph{aborted pre-burst}}, {\color{red} \emph{trigger}}, {\color{brun} \emph{successful pre-burst}},  and {\color{green} \emph{burst}}.}
\label{FigActivityAllDef}
\end{figure}

\subsection{Definitions}\label{definition}

Figure \ref{FigActivityAllDef} represents a part of a simulation and leads us to define sequences in the simulations. This section is roughly the same than \cite[Section 2.3]{Leadership}. Precisely we will define mathematically what we call a burst and a leader and we will also explain what we call mutual information. In order to make the present paper comparable to \cite{Leadership}, we adhere strictly to the same definitions. So the reader who red \cite{Leadership} or knows these definitions can skip this part and move directly to the next section.

\subsubsection{Definition of bursts and their triggers}\label{BurstDefinition}

First we divide all the spikes of our list into four classes: \emph{burst}, \emph{successful pre-burst}, \emph{aborted pre-burst}, \emph{isolated spike}, and each spike belonging to one and only one class. The precise definitions are given below, but basically a spike is in a burst if it is in a group of many spikes that follow each other closely. If a spike is not in a burst it can be in a successful pre-burst or aborted pre-burst. But it must be anyway is in a sequence of spikes that are close enough in time so that communication between them is still possible. The distinction between successful or aborted pre-burst depends on whether the spike sequence is eventually followed by a burst or not. Finally, all other spikes are isolated.

More precisely, we use three parameters, $n_\b$, $\delta t_\b$, and $\delta t_\p$. We first look for interspike gaps of length at least $\delta t_\p$, and we divide the set of all spikes into disjoint subsets  $\RR_k$ of consecutive spikes, where $k$ is a running index. The $\RR_k$ have the property that if (the spike) $i$ and $i+1$ are in $\RR_k$ then $t_{i+1}-t_i \le \delta t_\p$, while if $i\in\RR_k$ and $i+1\in \RR_{k+1}$ then $t_{i+1}-t_{i}>\delta t_\p$. The rationale is that if the interspike gap is so large that the spike $i+1$ is due to noise and not to the axon signaling of the precedent (in time) spike $i$ (\ie $\ t_{i+1}-t_{i}>\delta t_\p$ then the spikes belong to separate subsets).

We now want to further divide each $\RR_k$ into the burst itself, characterized by a high density of spikes, and its precursor which immediately precedes it in time but has a lower density. Each of the $\RR_k$ contains at least one spike, but may contain many. In each $\RR_k$ we first look for a spike that is followed by at least $n_\b-1$ spikes in $\RR_k$ within a lapse of time $\delta t_\b$. Denoting this spike's index by $i_*=i_*(k)$, this is the first index in $\RR_k$ with the property:
\begin{equation}\label{conditionSim}
t_{i_*+n_\b-1}-t_{i_*}<\delta t_\b~.
\end{equation}
If the condition \eqref{conditionSim} is met then we subdivide $\RR_k$ into two disjoint sets at $i_*$: $\RR_k=\PP_k\cup \BB_k$ ($\PP_k$ can be empty). The index $i_*$ up to the last index in $\RR_k$ make up the burst $\BB_k$.

If the condition \eqref{conditionSim} is never met in $\RR_k$, then the set $\RR_k$ is not subdivided. This is called an \emph{aborted pre-burst} if it has more than one spike and is an \emph{isolated spike} otherwise.

For these $k$ where such an $i_*$ can be found, we check if it is also the first spike in $\RR_k$. If so, this burst is an \emph{immediate burst} that has no successful pre-burst, so that $\RR_k=\BB_k$. For all the other bursts we divide $\RR_k=\PP_k\cup \BB_k$ where $\PP_k$ are the index $i\in\RR_k$ with $i < i_*$, and $\BB_k$ are the others. The letter $\PP$ refers to successful pre-bursts and $\BB$ refers to bursts.

Between bursts the neural activity is much lower, and we label these periods as \emph{quiet}. Each quiet period, denoted $\QQ_b$, for a running index $b$, is actually a concatenation of consecutive $\RR_k$'s that did not contain a burst in the earlier stage, or else an empty set. We now renumber the set of index as follows
\begin{equation*}
\{1,\dots,M\} = \QQ_1 \cup \PP_1 \cup \BB_1\cup \QQ_2\cup \PP_2\cup \BB_2 \cup
\dots~,
\end{equation*}
with $\BB_b$ the $b$'th burst, $\PP_b$ the corresponding successful pre-burst (if it exists else $\PP_b$ is empty) and $\QQ_b$ the corresponding quiet phase. We define the \emph{trigger} of burst $b$ as the neuron $n_{i_b}$, where $i_b$ is the number of the first spike in $\PP_b$ (resp.~$\BB_b$ if $\PP_b$ is empty like for an immediate burst).

\begin{table}[ht]
\begin{center}
\begin{tabular}{|c|c|c|c|c|}
   \hline
   $N$ & Analyzed neurons & $n_\b$ & $\delta t_\b$ [ms] & $\delta t_{\p}$ \\
   \hline
   \multirow{2}{*}{$900$} & $60$ & $20$ & \multirow{2}{*}{$15$} & \multirow{2}{*}{$\frac{2}{3}\frac{\sqrt{N}}{v}$} \\
   & $900$ & $>100$ & & \\
   \hline
\end{tabular}
\end{center}
\caption[Burst detection parameters]{Burst detection parameters.} \label{tab:BurstParamTableSim}
\end{table}

The parameters $n_\b$, $\delta t_\b$, and $\delta t_\p$ that we used for $N=900$ are shown in Table \ref{tab:BurstParamTableSim}. Note that these parameters can easily be adapted for other values of the number of neurons $N$. Remark also that the time parameters $\delta t_\p$ and $\delta t_\b$ have to be smaller than the characteristic time of the membrane $\tau$. The rationale for these choices are: Firstly, to cross the distance of the biggest possible axon $\frac{2}{3}\sqrt{N}$, the signal, at the velocity $v$, needs a time of $\frac{2}{3}\frac{\sqrt{N}}{v}$. This computation gives us the order of $\delta t_{\p}$. This means that only the singnaling velocity is taken into account to compute $\delta t_{\p}$. Secondly, even in a burst, the neural activity is not regular, so to balance this effect we need $\delta t_\b$ to be high enough. Finally, for $n_\b$ we want it high enough to guarantee that a burst includes many neurons.

\subsubsection{Leaders}\label{LeaderDefinition}

For every successful pre-burst, aborted pre-burst or immediate burst, the neuron which fires first is called the \emph{trigger}. Since we are interested in their special r\^ole, we first need to make sure that triggers are not just neurons with high neural activity, which are statistically more often the first ones to fire. The following discussion will exhibit that leaders are over-proportionally more often triggers.

To qualify a neuron as a leader, we require that a trigger's probability to lead bursts should be significantly higher than its probability to fire. Let $M$ be the total number of bursts in a simulation. For each neuron $n$, we define $a_n$ as the number of times that neuron $n$ has spiked during the simulation. Now, let us consider a spike, and term by $q_n$ the probability\footnote{unbiased estimator of the considered probability\label{unbiasEstimP}} that neuron $n$ has fired that spike: $q_n=a_n/\sum_{n'} a_{n'}$. The probability for the neuron $n$ to be a spurious, or random trigger $F$ times in $M$ bursts is given approximately\footnote{It would be exactly the binomial distribution if all the spikes were independent. All the spikes are certainly not independent because neurons are connected. So the binomial distribution is not the correct one but gives a good enough approximation (see \cite{MaThese}).} by the binomial distribution $P_n(F)$:
\begin{equation}\label{LeaderAlpha}
\begin{pmatrix}M\\F\end{pmatrix}q_n^{F}(1-q_n)^{M-F}~.
\end{equation}
In the limit of large $M$ and reasonable $q_n$ this distribution is approximated by a Gaussian of mean $Mq_n$ and variance $Mq_n(1-q_n)$. On the other hand, we denote by $f_n$ the actual number of bursts that neuron $n$ leads (note that $\sum_n {f_n} = M$).

Thus we have a scale on which to test triggering. We define $\alpha_n$, a ``leadership score'', and decide that a neuron is a \emph{leader} if it scores at least $3$ standard deviations above the natural expectation value. The criterion for leadership of neuron $n$ is therefore
\begin{equation}\label{LeadershipScore}
\alpha_n=\frac{f_{n}-Mq_{n}}{\sqrt{Mq_{n}(1-q_{n})}}  ~>~ 3~.
\end{equation}
Because we expect $\alpha\sim \mathcal{N}(0,1)$ to be Gaussian, \eqref{LeadershipScore} corresponds to a p-value of $0.001$.

\subsubsection{Mutual Information}\label{MutualDefinition}

We estimate the mutual information of two neurons (pair of neurons) in a series of time intervals based on empirical probabilities, according to \cite{Email}. Using the division (of time) into aborted pre-burst, successful pre-burst and burst intervals, we define a firing event in a given interval as $h=0,1$, where $0$ stands for no spike and $1$ stands for at least one spike. We denote by $N_n(h)$ the number of times neuron $n$ had an event $h$ in a series of $N_{\rm int}$ intervals\footnote{Note that $N_{\rm int}=M$ for the burst intervals and $N_{\rm int}\leq M$ for the successful pre-burst intervals.}. The probability assigned to the event $h$ is thus$^{\ref{unbiasEstimP}}$ $p_n(h)=N_n(h)/N_{\rm int}$. The number of joint events in which neuron $n$ had an event $h_1$ and neuron $n'$ had an event $h_2$ is given by $N_{n,n'}(h_1,h_2)$. The joint probability of events $h_1,h_2$ for neurons $n$, $n'$ is then given by$^{\ref{unbiasEstimP}}$ $p_{n,n'}(h_1,h_2)=N_{n,n'}(h_1,h_2)/N_{\rm int}$. The mutual information between two neurons is then given by
\begin{equation*}
I_{n,n'}=\displaystyle\sum_{h_1,h_2\in\{0,1\}}p_{n,n'}(h_1,h_2)\cdot
\log\displaystyle\frac{p_{n,n'}(h_1,h_2)}{p_n(h_1)p_{n'}(h_2)}~.
\end{equation*}
Note that the mutual information between two neurons is $0$ if the two neurons fire independently and is $\ln 2$ if the two neurons fire always together (see \cite{MultiInf}). These two extreme values give us the variation scale of the mutual information for a pair of neurons.

\section{Results}\label{results}

In this section we explain our results. We prove that our simulations have
stable leaders and we identify the topological properties of these leaders. An analysis of the first to fire neurons exhibits that the leader properties we find are the correct ones.

\subsection{About the parameters and some statistics}\label{AboutParam}

The next step consists in analyzing our simulations knowing that our results will be more consistent if we span as many parameters as possible. However, our goal is not to obtain a transition phase diagram \cite{WulfAdaptExpo}, so we are not going to vary all parameters (most of the them were fixed in Section \ref{parameters} through Table \ref{tab:Param}). Up to now we stayed unclear about some parameters, including
dynamical parameters like $\w$, $\sigma$ and $\omega$. So we still have to find (and fix) a range for the parameters $\Delta V^{*}$, $\w$, $\sigma$ and $\omega$ in a way to obtain ``reasonable'' simulations\footnote{Here by ``reasonable'' simulations, we mean to observe a neural activity more or less like the one in Figure \ref{FigActivityAllDef} but without a precise predicting spike timing \cite{WulfAdaptMain,WulfPredict}.} which involves distinguishing bursts, successful pre-bursts and quiet periods. One of the best ways to do that mathematically is to look at the time interspike distribution and compare with the experimental one (data not shown, see \cite{MaThese}). The analysis of the membrane potential profile and the time length distribution are also possible in the simulations (see \cite{MaThese}).

\subsection{Existence of leaders}\label{leadersexistence}

\begin{figure}[ht]
 \centering
  \subfloat[The leadership score when we look at only $60$ neurons like in \cite{Leadership}. We see that we have leaders and even very good ones. We also see that the average leadership score {\color{red}$\langle\alpha\rangle$} is about $0$. Note that in this graph, each point represents $600$ seconds of simulation that contains more than $2500$ bursts. Remark that the {\color{brun}best leader} among $60$ neurons is not necessarily a leader while we consider all the neurons (see Figure \ref{FigLeaderScoreAll}). In the restricted case, the neurons detected as leaders can be first to fire neurons (see Section \ref{firsttofire}).]{
\begin{psfrags}\label{FigLeaderScore60}%
\psfrag{Leadership score = f(time)}[c][c][1]{\scriptsize Leadership score as a function of time}%
\psfrag{60 neurons}[c][b][1]{\tiny $60$ neurons}%
\psfrag{Leadership score [-]}[c][t][1]{\tiny Leadership score $\left(\alpha_n\right)$}%
\psfrag{Time [minutes]}[c][b][1]{\tiny Time [minutes]}%
\psfrag{Neuron 655}[l][c][1]{\tiny Neuron 655}%
\psfrag{Neuron 643}[l][c][1]{\tiny {\color{blue}Neuron 643}}%
\psfrag{Neuron 634}[l][c][1]{\tiny {\color{brun}Neuron 634}}%
\psfrag{Neuron 190}[l][c][1]{\tiny Neuron 190}%
\psfrag{Neuron 187}[l][c][1]{\tiny Neuron 187}%
\psfrag{Average alpha (60 neurons)}[c][c][1]{\tiny {\color{red}$\langle\alpha\rangle$} ($60$ neurons)}%
\includegraphics[width=0.36\textwidth,angle=-90]{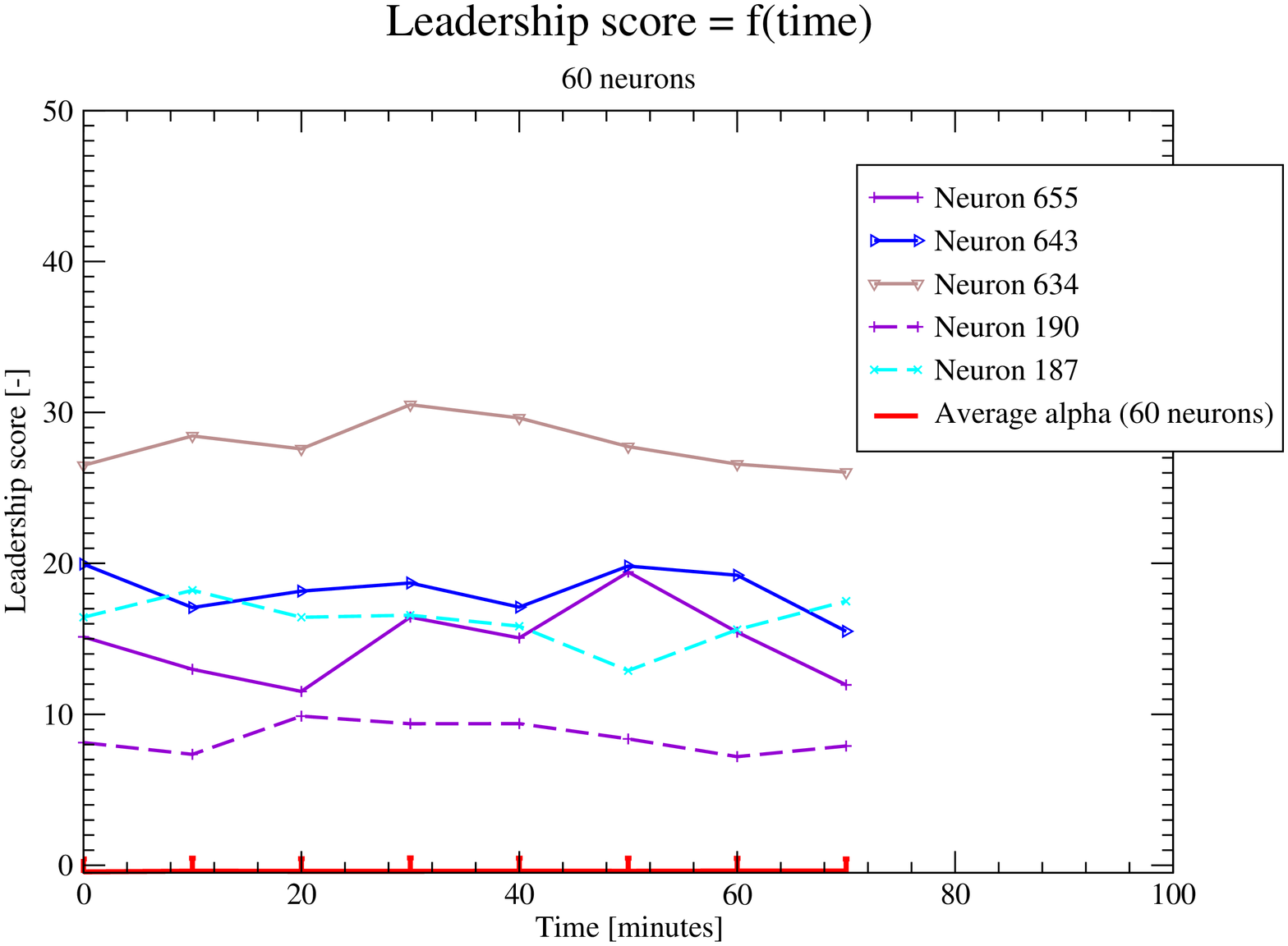}%
\end{psfrags}}\hfill%
  \subfloat[The leadership score when we consider all the $900$ neurons. We can say that these neurons are the ``true'' best leaders of the simulation. Note also that, luckily, {\color{blue}one} of the best leaders was in the $60$ selected neurons. But even when it is not the case, when we consider only $60$ neurons, some neurons show very good leadership scores. On this graph, each point represents about $2000$ seconds of simulation that contains more than $8800$ bursts.]%
{\begin{psfrags}\label{FigLeaderScoreAll}%
\psfrag{Leadership score = f(time)}[c][c][1]{\scriptsize Leadership score as a function of time}%
\psfrag{All 900 neurons}[c][b][1]{\tiny All the $900$ neurons}%
\psfrag{Leadership score [-]}[c][t][1]{\tiny Leadership score $\left(\alpha_n\right)$}%
\psfrag{Time [minutes]}[c][b][1]{\tiny Time [minutes]}%
\psfrag{Neuron 834}[l][c][1]{\tiny Neuron 834}%
\psfrag{Neuron 634}[l][c][1]{\tiny {\color{brun}Neuron 634}}%
\psfrag{Neuron 643}[l][c][1]{\tiny {\color{blue}Neuron 643}}%
\psfrag{Neuron 437}[l][c][1]{\tiny Neuron 437}%
\psfrag{Neuron 427}[l][c][1]{\tiny Neuron 427}%
\psfrag{Average alpha (900 neurons)}[c][c][1]{\tiny {\color{red}$\langle\alpha\rangle$} ($900$ neurons)}%
\includegraphics[width=0.36\textwidth,angle=-90]{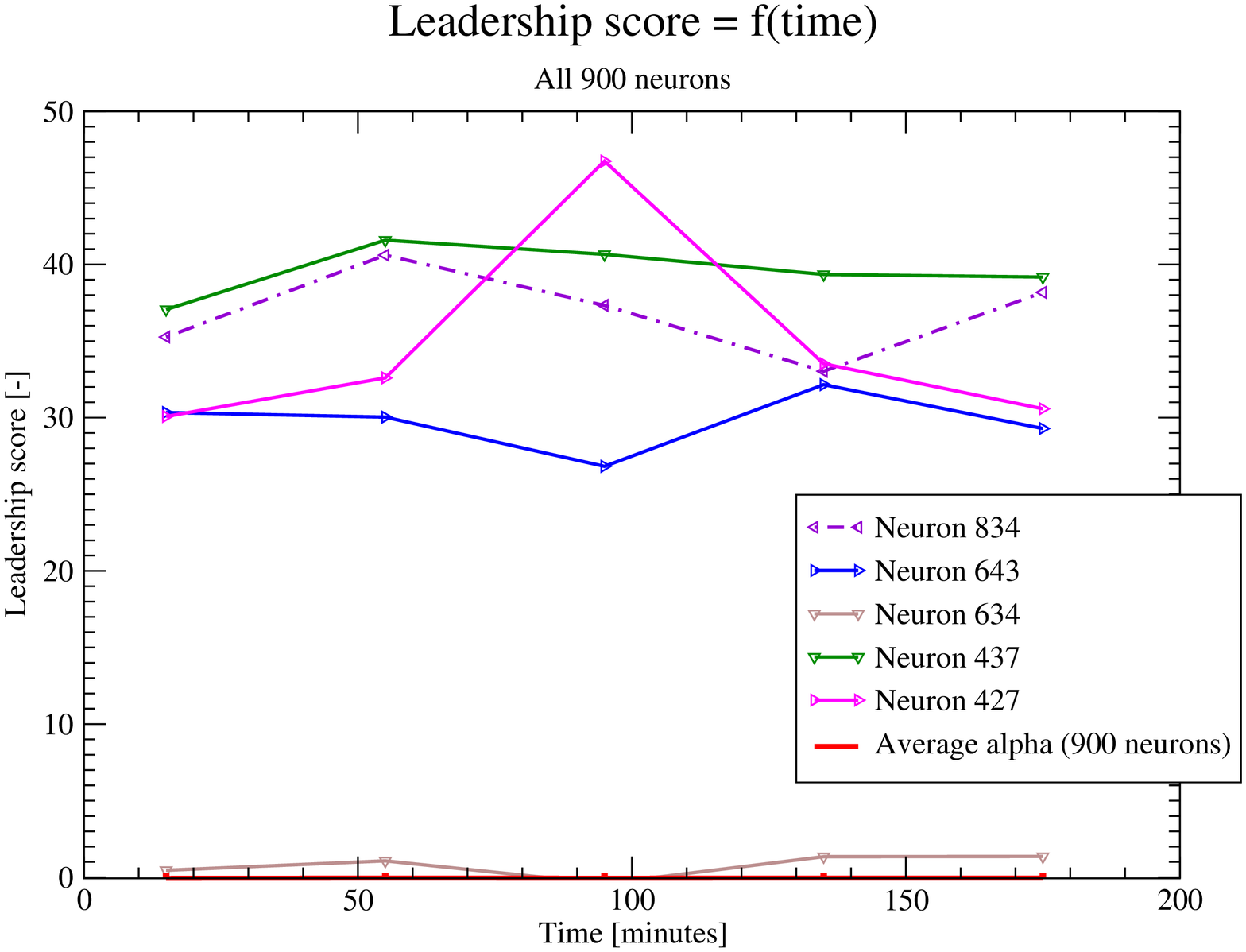}%
\end{psfrags}}%
\caption[Leadership score of a few neurons for a simulation]{Leadership score of a few neurons for the simulation shown in Figures \ref{FigActivityAllDef}. Note that the leadership score depends on the number of neurons considered for the analysis (see neuron {\color{brun}634}, {\color{blue}643} and Section \ref{firsttofire}).}\label{FigLeaderScore}%
\end{figure}

In Section \ref{LeaderDefinition}, we defined, for every neuron $n$ a leadership score $\alpha_n$ (see \eqref{LeadershipScore}) that compares the triggering frequency to the leadership expected frequency (compute with respect to the neural activity frequency). Figures \ref{FigLeaderScore} show the leadership score as a function of the time for a few neurons. In these figures we see that our simulations have leaders. Note that this result is very robust. We mean that the choice of the used parameters in simulations is not important. As soon as we use ``reasonable''parameters, our simulations have leaders. This is still true when we
only look at a few neurons instead of all them (Figure \ref{FigLeaderScore60}). Even more, we can change the axons length distribution and the simulations still have stable leaders. Note also that even a network with a uniform probability of connections\footnote{This means a network that is not build with the kind of geometry showed in Figure \ref{FigGeometry}.} shows bursts and leaders but, in this case, we do not observe any spatial locality.

In our simulations, with $N>400$ about $5\%$ to $10\%$ of the neurons have a mean leadership score \eqref{LeadershipScore} bigger than the threshold $3$ when we consider all the neurons. While considering only $60$ neurons, about $15\%$ neurons have a mean leadership score bigger than the threshold $3$. Also note that even if the leadership score can fluctuate during the simulations, its mean value stays stable (\ie  its standard deviation is small). Of course, the leader ratio decreases when the mean noise $\sigma$ increases. This fact is logical because in a too noisy neural activity it becomes more difficult to identify the trigger of each burst. Note also that it seems that when  $\Delta V^{*}$ (standard deviation of the membrane potential firing threshold) is between $0.05$ and $0.1$ the leader ratio seems maximal (data not shown).

In the experiments \cite{Leadership}, one electrode captures in general the spike signal of several neurons \cite[Figure 1a]{Leadership}. But, in our simulations, up to this point we considered each neuron separately, as if one electrode measures only one neuron. Remark that the leadership score of a group of neurons also reach the value $3$ for some groups \cite{MaThese}.

At this point, our simulations reproduce basically all the findings of \cite{Leadership} (presence of leaders, signature of bursts, force of leaders,\ldots) see \cite{MaThese}. The advantage of simulations is that we have access to a lot of details that we did not have access to in the experiments and we can analyze the topology of the network to find out the properties of leaders.

\subsection{Leader properties}\label{leadersproperties}

\subsubsection{Some facts about the leader properties}\label{leaderfacts}

\begin{figure}[ht]
 \centering
  \subfloat[The membrane potential firing threshold $(V^*_n)$ compared to the leadership score $(\alpha_n)$.
  We see that good leaders have a low membrane potential firing threshold $\left(\langle V^{*}\rangle=1\right)$.
  Parameters: $N=900$, $\r=0.85$, $L=7$, $r=0.2$, $\Delta V^{*}=0.1$, $v=5$ $\frac{1}{\textrm{ms}}$,
  $\w=0.1$, $\tau=100$ ms, $\sigma=0.008$, $\omega=0.001$ s, $n_\b=100$, $\delta t_\b=15$ ms and
$\delta t_{\p}=5$ ms.]{
\begin{psfrags}\label{FigLeaderScorePot}%
\psfrag{Membrane potential firing threshold VS Mean leadership score}[c][c][1]{\scriptsize Firing threshold compared to leadership score}%
\psfrag{Mean leadership score}[c][b][1]{ \tiny Leadership score $\left(\alpha_n\right)$}%
\psfrag{Membrane potential firing threshold}[c][t][1]{ \tiny Membrane potential firing threshold $\left(V_{n}^{*}\right)$}%
\includegraphics[width=0.36\textwidth,angle=-90]{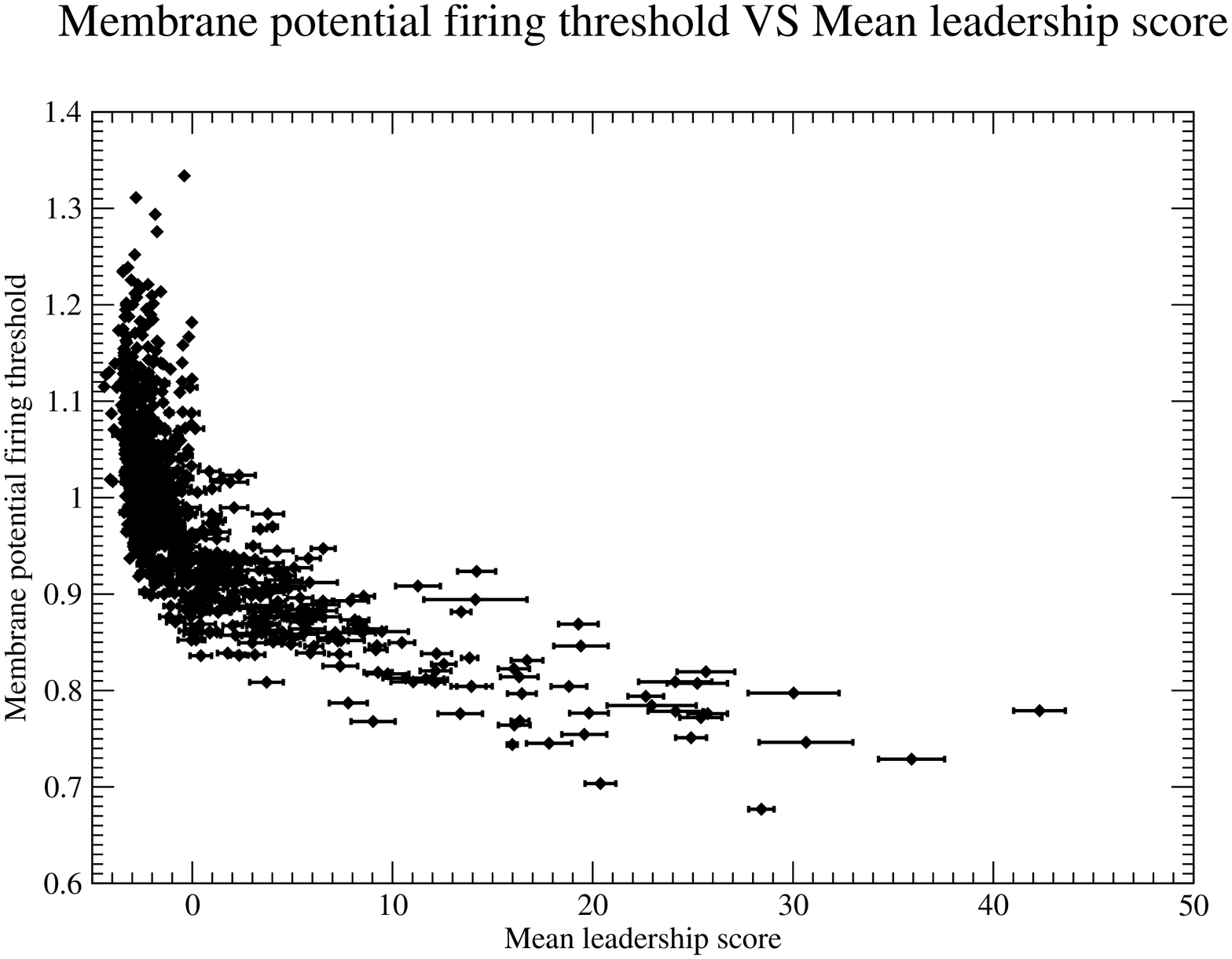}%
\end{psfrags}}\hfill%\hspace{0.5cm}
  \subfloat[The triggering frequency $\left(f_n/M\right)$ compared to the leadership score $(\alpha_n)$.
  These two things are proportional.
  Parameters: $N=900$, $\r=0.85$, $L=7$, $r=0.2$, $\Delta V^{*}=0.05$, $v=5$ $\frac{1}{\textrm{ms}}$,
  $\w=0.1$, $\tau=100$ ms, $\sigma=0.008$, $\omega=0.001$ s, $n_\b=100$, $\delta t_\b=15$ ms and
$\delta t_{\p}=5$ ms.]%
{\begin{psfrags}\label{FigLeaderScoreBoss}%
\psfrag{Triggering score VS Mean leadership score}[c][c][1]{\scriptsize Triggering frequency compared to leadership score}%
\psfrag{Triggering score}[c][t][1]{ \tiny Triggering frequency $\frac{f_n}{M}$}%
\psfrag{Mean leadership score}[c][b][1]{\tiny Leadership score $\left(\alpha_n\right)$}%
\includegraphics[width=0.36\textwidth,angle=-90]{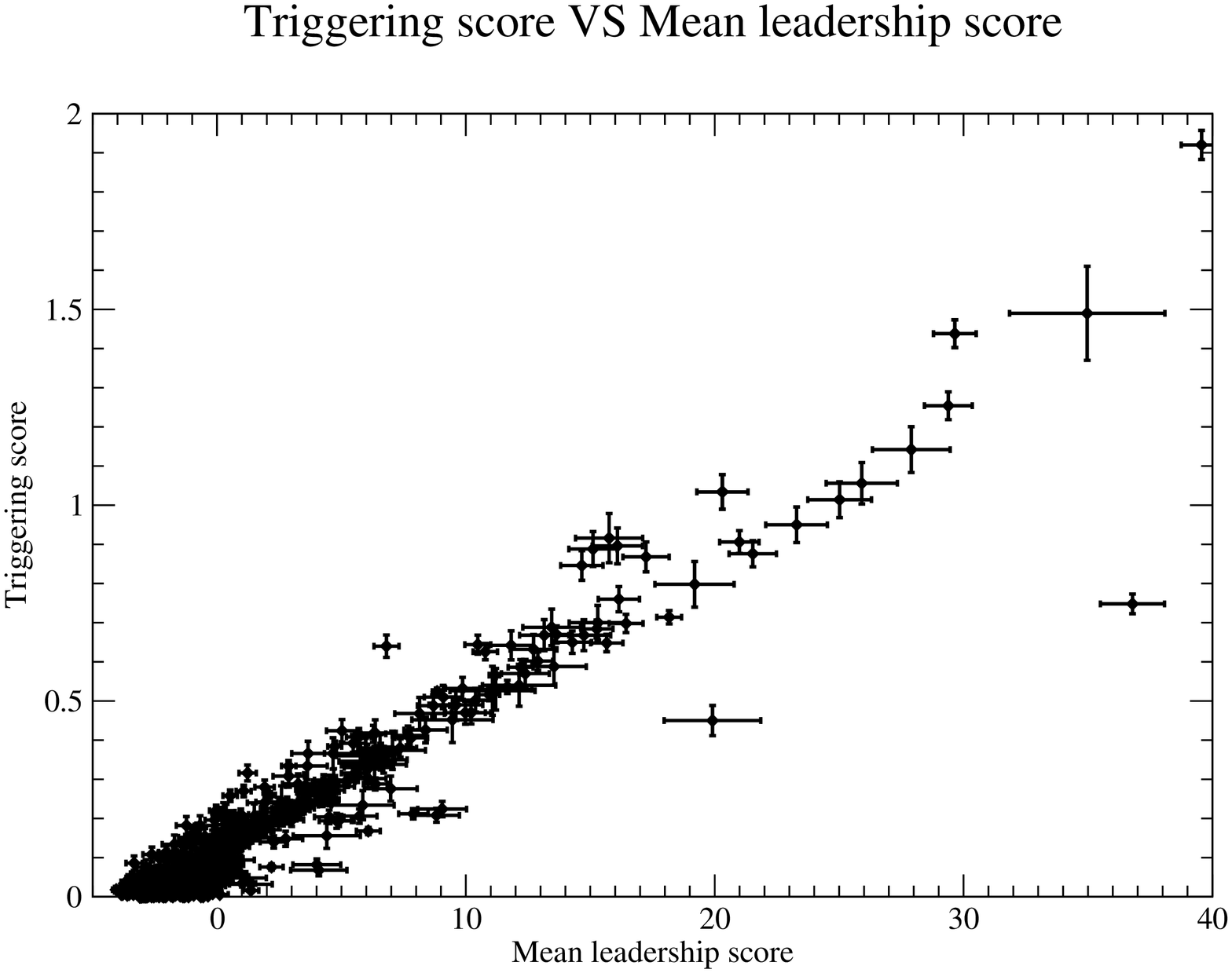}%
\end{psfrags}}%
\caption[Some properties of leaders for simulations]{Some properties of leaders for simulations with parameters close to the one we showed in Figure \ref{FigActivityAllDef}. Note that each point in these figures represents an average of the results for $5$ simulations using the same neural network and the same parameters. Each simulation contains more than $10000$ bursts and the simulated time was longer than $25$ minutes.}\label{FigLeaderProperties}%
\end{figure}

In Figures \ref{FigLeaderProperties} we see clearly that the
leaders seem to have some special properties compared to the other
neurons. More precisely, Figure \ref{FigLeaderScorePot} shows that
good leaders typically have a low membrane potential firing
threshold. This fact exhibits that to be a good leader, a neuron
must have the property of firing early. But, this property is not
the only one a neuron needs to have to be a leader. Indeed the
best leader in Figure \ref{FigLeaderScorePot} is not the neuron
with the lowest membrane potential firing threshold. Thus, another
property of the leaders is easy to highlight: the leaders are
excitatory neurons\footnote{Remark that in the special case where
$\omega\ll\tau$, we rarely observed a few inhibitory neurons with a
leadership score bigger than $3$. This can happen only when these
inhibitory neurons have a very low membrane potential firing
threshold, a few excitatory fathers and a lot of inhibitory fathers.
The reason is as follow, because $\omega\ll\tau$, the network gets
synchronized after the first burst and these particular inhibitors
have the faculty to fire even before the true trigger of the burst.
Beside the type of their fathers implies a very low neural
activity, so their leadership scores are potentially good. Note that
the few top best leaders are always excitatory neurons.} (data not shown).
In Figure \ref{FigLeaderScoreBoss}, we see that, like in \cite{Leadership}, the better a leader is, the more
bursts it triggers.

\begin{figure}[ht]
\begin{center}
\parbox{0.47\textwidth}{%
\begin{psfrags}%
\psfrag{Mean leadership score VS Activity ratio}[c][c][1]{\scriptsize Leadership score compared to the neural activity ratio}%
\psfrag{Mean leadership score}[c][t][1]{\tiny Leadership score $\left(\alpha_n\right)$}%
\psfrag{Activity ratio}[c][b][1]{\tiny Neural activity ratio $\left(q_n\right)$}%
{\includegraphics[width=0.36\textwidth, angle=-90]{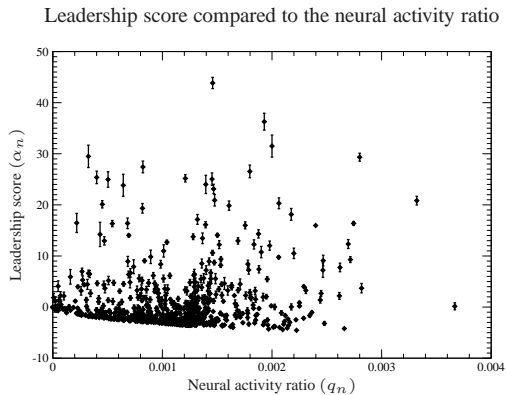}}%
\end{psfrags}
}\parbox{0.53\textwidth}%
{\caption[The leadership score compared to the neural activity ratio]{ The leadership score $\left(\alpha\right)$ compared to the neural activity ratio $\left(q\right)$. Note that, as before, each point in this figure represents an average of the results for $5$ simulations using the
same neural network. Parameters: $N=900$, $\r=0.85$, $L=7$, $r=0.2$, $\Delta V^{*}=0.1$, $v=5$ $\frac{1}{\textrm{ms}}$, $\tau=100$ ms, $\w=0.1$, $\sigma=0.008$, $\omega=0.001$ s, $n_\b=100$, $\delta t_\b=15$ ms and $\delta
t_{\p}=5$ ms.}\label{FigLedActiv}}
\end{center}
\end{figure}

We can also have a look at the leadership score compared to the
neural activity ratio of the neurons shown in Figure
\ref{FigLedActiv}. Firstly, remember that to compute the
leadership score we use the neural activity ratio $q$ (see
\eqref{LeadershipScore}). Secondly, look at Figure
\ref{FigLedActiv} and remark that there are as much leaders with a
high neural activity ratio as leaders with a low neural activity ratio. So,
clearly and as expected, for a given neuron there is no link
between having a high leadership score and the neural activity ratio,
even if a good leader triggers a lot of bursts (see Figure
\ref{FigLeaderScoreBoss}).

Before trying to find more leader properties, we add more
remarks about Figure \ref{FigLedActiv}. We note that there
exist some neurons with a very high positive leadership score
($\alpha>10$) but there is no neuron with a very low leadership
score ($\alpha<-10$). We also note that if the standard
deviation of the membrane potential firing threshold $\Delta V^{*}$ is $0$
then most of the neurons have a neural activity ratio near
$\bar{q}=\frac{1}{N}$ (in Figure \ref{FigLedActiv}
$\bar{q}=\frac{1}{900}\cong0.11\%$ as expected).

Unfortunately, the other relations between a property and the
leadership score that we tried to highlight did not exhibit an as
good correlation as the relations showed in Figures
\ref{FigLeaderProperties}. We also note another
important fact: In the special case where we use a network with
every membrane potential firing threshold fixed to $\langle V^{*}\rangle$ (\ie $\
\Delta V^{*}=0$) or/and without inhibitory neurons (\ie $\ r=0$), our
simulations still have stable leaders. These two facts lead us to
the following hypothesis: the properties of leaders are a non
trivial combination of properties.

\begin{figure}[ht]
\begin{center}
\parbox{0.47\textwidth}{%
\begin{psfrags}%
\psfrag{Leadership score compare}[c][c][1]{\scriptsize Leadership score comparison}%
\psfrag{Leadership score w 0.1 N 0.014 dt 0.002}[c][t][1]{ \tiny Leadership score $\alpha_n$($\w=0.1$, $\sigma=0.014$, $\omega=0.002$ s)}%
\psfrag{Leadership score w 0.1 N 0.0081 dt 0.001}[c][b][1]{ \tiny Leadership score $\alpha_n$($\w=0.1$, $\sigma=0.0081$, $\omega=0.001$ s)}%
\includegraphics[width=0.36\textwidth, angle=-90]{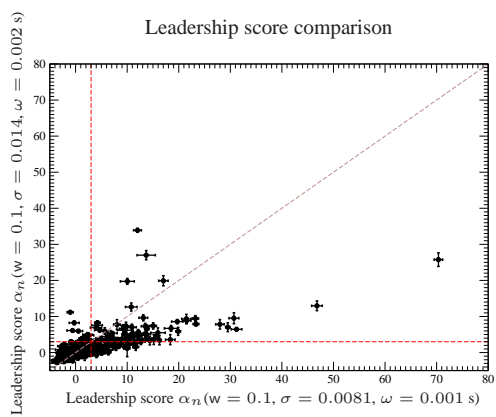}%
\end{psfrags}
}\parbox{0.53\textwidth}%
{\caption[The leadership score of two different simulations]{The leadership score of two different simulations
using the same neural network. The {\color{red} red dashed lines} separate leaders and non-leaders for both simulations. We observe that the best leader is not necessarily the same and some neurons are leaders with a set of parameters but are not leaders with the other set. We also observe that most of the time a good leader
continues being a good leader in both simulations. Like for
Figures \ref{FigLeaderProperties}, the error bars are computed by averaging over $5$ simulations using the same network and dynamical parameters $\w$, $\sigma$ and $\omega$. Common parameters: $N=900$,
$\r=0.85$, $L=7$, $r=0.2$, $\Delta V^{*}=0.1$, $v=5$
$\frac{1}{\textrm{ms}}$, $\tau=100$ ms, $n_\b=100$, $\delta t_\b=15$ ms and $\delta
t_{\p}=5$ ms.}\label{FigLeaderRunCompare}}
\end{center}
\end{figure}

To find these properties, we
perform the following experiment: We take a given
network\footnote{This means that we choose $N$, $\r$, $L$, $p$,
$\Delta V^{*}$, $r$ and build a realization of a network (\ie $\ W$ and
$V_n^*$ are known for all $n$).} and make several simulations with
this network but change some of the dynamical parameters ($\tau$,
$v$, $\w$, $\sigma$ and $\omega$) in a way of keeping a
reasonable\footnote{Here by a reasonable simulation, we mean that
we want to observe quiet periods, successful pre-bursts and
bursts in a way to find leaders. Note that this range of
reasonable parameters is quite large and the relation between the
dynamical parameters is non trivial, see remark in Section
\ref{parameters} and Section \ref{AboutParam}.} simulation. Then,
we compute the leadership score of each neuron and we do this for each
simulation. As a result, we find that the best leader is not
always the same even if most of the time a leader stays
leader. Figure \ref{FigLeaderRunCompare} illustrates this fact and
this forces us to admit that the leader properties depend on the
dynamical parameters. Nevertheless, we can probably find a
trend of the properties of a typical leader by doing a linear
least squares analysis for all our simulations. Indeed, in Figure
\ref{FigLeaderRunCompare} more than $90\%$ of the leaders keep on being leaders and non leaders
keep on being non leaders under the change of dynamical
parameters.

Before going through this linear least squares
analysis, note that in the experiments related in \cite{Leadership}, the leaders change during the days spent \emph{in vitro} with a typical life time of about one day (see \cite[Figure 3a]{Leadership}). While, in the simulations, the leaders stay
the same in a given network if we do not change the dynamical
parameters of the simulation. So, one of the interesting facts in
the simulations is that to change the leaders we do not need to
change the network, some changes in the dynamical parameters are
enough (we consider the synaptic weight $\w$ as a dynamical parameter).

\subsubsection{Finding the leader properties}\label{meanleaderproperties}

To find out the leader properties we do the following linear least square analysis. First, we build a vector $\vec\alpha=\left\lbrace\alpha_n\right\rbrace_{n=1,\ldots,N}$ that contains the leadership score $\alpha_n$ of each neuron $n$ of a given simulation. Then we construct a matrix $P=\left\lbrace P_{ni}\right\rbrace_{1\leq n\leq N,\ 1\leq i\leq6}$ that contains all the neuron properties (excitatory or inhibitory neuron and membrane potential firing threshold) and all the connection properties at the first order\footnote{We call connection properties at the first order all the direct connections of the neuron in the network contained in the matrix of synaptic weights $W$ (this means the number of sons and fathers (see Section \ref{strategy})).}. More precisely: $P_{n1}=1$ if the neuron $n$ is an excitatory neuron and $-1$ otherwise. $P_{n2}=\frac{V_n^*-\langle V^{*}\rangle}{\Delta V^{*}}$ if $\Delta V^{*}>0$ and $0$ otherwise. $P_{n3}$ is number of excitatory neurons connected to the axon of neuron $n$ (this means the number of excitatory sons of the neuron $n$, see Section \ref{strategy}) and $P_{n4}$ the number of inhibitory sons. Identically, $P_{n5}$
is the number of excitatory fathers of the neuron $n$ and $P_{n6}$
the number of inhibitory fathers.

Our linear least square method, like in \cite{LinAlg}, consists in
finding a vector  $\vec x=\left\lbrace
x_i\right\rbrace_{i=1,\ldots,6}$ such that the euclidean norm
\begin{equation}
\label{linear}
\left\|P\vec x-\vec\alpha\right\|
\end{equation}
is minimized. This particular $\vec x$ gives us the relations (at
the linear order) between the first order properties\footnote{We
call first order properties all the neuron properties and all the
connection properties at the first order.} of a neuron and its
leadership score.

By repeating the calculation \eqref{linear} for each of our simulations and averaging over
all our simulations, we obtain the typical solution $\vec
x=\left(1.22,\ -2.33,\ 0.13,\ -0.19, \ -0.13,\ 0\right)^T$. This
means that a good prediction $p_n$ for the leadership score
$\alpha_n$ of the neuron $n$ in a network is given by
\minilab{predic}
\begin{equs}
p_n&=1.22\cdot P_{n1}-2.33\cdot P_{n2}+0.13\cdot P_{n3}-0.19\cdot P_{n4}-0.13\cdot P_{n5}\,,\ \ \ \ \ \ \ \label{predicdetail}\\
\textrm{if $\Delta V^{*}>0$: }\ \ \ p_n&=
\begin{array}{{l}}
1.22\cdot\textrm{excitatory}_n-2.33\cdot\frac{V_n^*-\langle V^{*}\rangle}{\Delta V^{*}}+0.13\cdot\textrm{\# excitatory sons}_n\\[2mm]-0.19\cdot\textrm{\# inhibitory sons}_n-0.13\cdot\textrm{\# excitatory fathers}_n
\end{array}\ \ \ \ \ \ \ \ \label{predicExplicit}
\end{equs}
where $\left(\textrm{excitatory}_n=\right)P_{n1}=1$ if the neuron $n$ is an excitatory neuron and $-1$
otherwise, $P_{n2}=\frac{V_n^*-\langle V^{*}\rangle}{\Delta V^{*}}$ if $\Delta V^{*}>0$ and $0$
otherwise, $P_{n3}\left(=\textrm{\# excitatory sons}_n\right)$ is the number of excitatory sons (of neuron $n$),
$P_{n4}\left(=\textrm{\# inhibitory sons}_n\right)$ is the number of inhibitory sons and $P_{n5}\left(=\textrm{\# excitatory fathers}_n\right)$ is the number of excitatory fathers.

Note also that \eqref{predic} do not depend on the axons length distribution. Even more, by building a fully random network (without axons and dendrites, \ie no particular geometry) and fixing only the mean number of connections per neuron, we obtain the same formula.

Precisely, equations \eqref{predic} mean that a typical leader in our
simulations is an excitatory neuron, that it has a low membrane potential firing threshold and that it
has a lot of excitatory sons but a few inhibitory sons and a few excitatory fathers. Equations \eqref{predic} in this exact form are valid for a mean number of connections per neuron of about eleven $\left(\langle\textrm{\# connections}/\textrm{neuron}\rangle\cong11\right)$.

\begin{figure}[ht]
 \centering
  \subfloat[The leadership score $\left(\alpha_n\right)$ compared to the prediction $\left(p_n\right)$ for a particular simulation. The {\color{red} red dashed lines} separate leaders and non-leaders for both the simulation and the prediction. As usual, the error bars of each point in this figure are computed using $5$ simulations done with the same network and parameters.]{
\begin{psfrags}\label{FigPredict}%
\psfrag{Prediction VS Leadership score}[c][c][1]{\scriptsize Leadership score compared to prediction}%
\psfrag{Leadership score}[c][t][1]{ \tiny Leadership score $\left(\alpha_n\right)$}%
\psfrag{Prediction}[c][b][1]{ \tiny Prediction $\left(p_n\right)$}%
\includegraphics[width=0.36\textwidth, angle=-90]{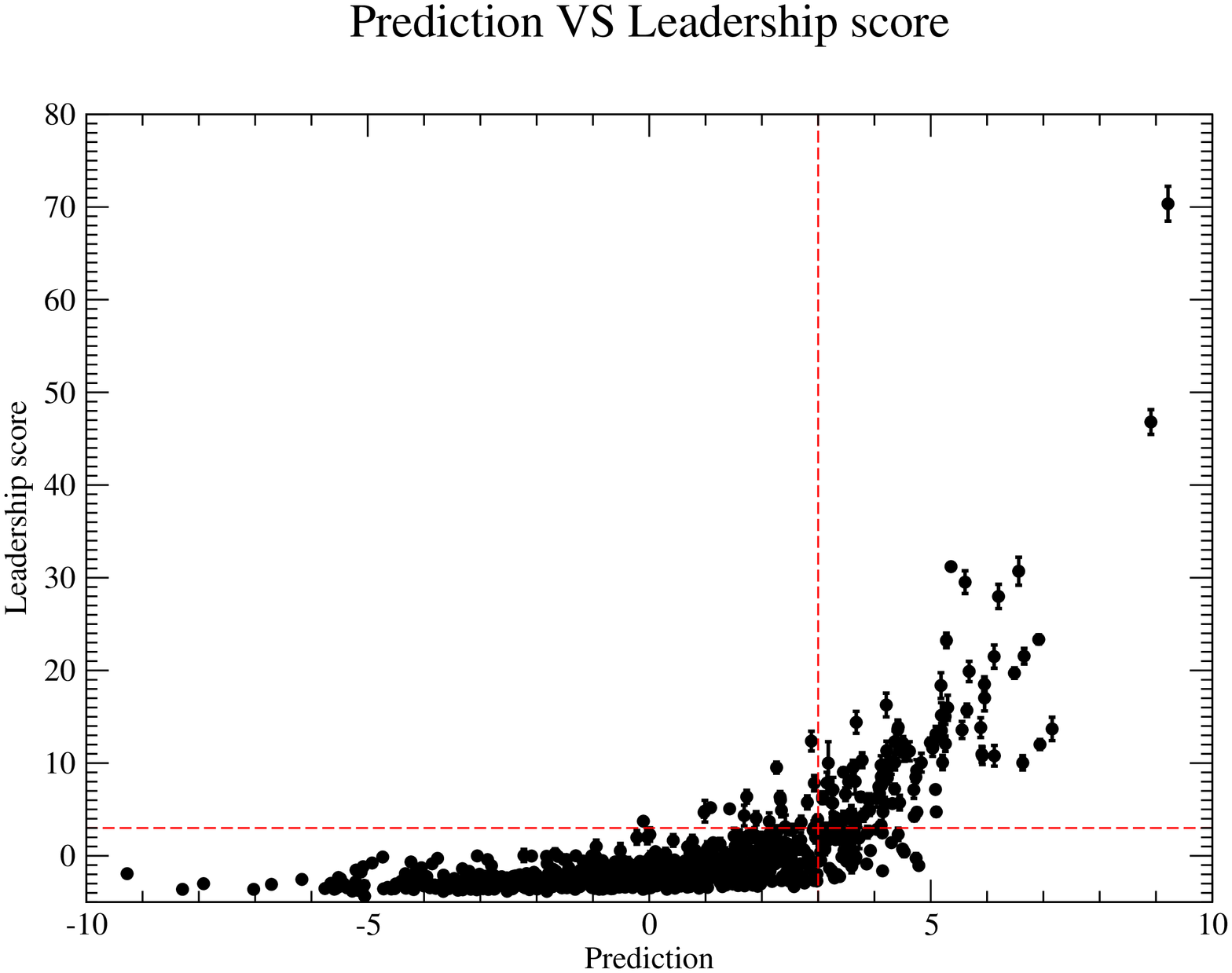}%
\end{psfrags}}\hfill%
  \subfloat[The prediction efficiency of Figure \ref{FigPredict}.
For each neuron $n$ we attach a number, named correct prediction$_n$, which is $1$ if the neuron $n$ is a leader $\alpha_n>3$ (respectively not a leader $\alpha_n\leq3$) in the simulation and its prediction $p_n>3$ (respectively $p_n\leq3$), otherwise: correct prediction$_n=0$. For each value of the leadership score $\alpha$, we compute the prediction efficiency $p_e(\alpha)=\frac{\sum_{n|\alpha_n<\alpha}\textrm{correct prediction}_n}{\sum_{n|\alpha_n<\alpha}1}$
  which is plot as a function of the leadership score $\alpha$.]%
{\begin{psfrags}\label{FigPredictEff}%
\psfrag{Prediction efficiency as a function of the leadership score}[c][c][1]{\scriptsize Prediction efficiency as a function of the leadership score}%
\psfrag{Prediction efficiency}[c][t][1]{ \tiny Prediction efficiency $p_e(\alpha)$}%
\psfrag{Leadership score}[c][b][1]{ \tiny Leadership score $\left(\alpha\right)$}%
\includegraphics[width=0.36\textwidth,angle=-90]{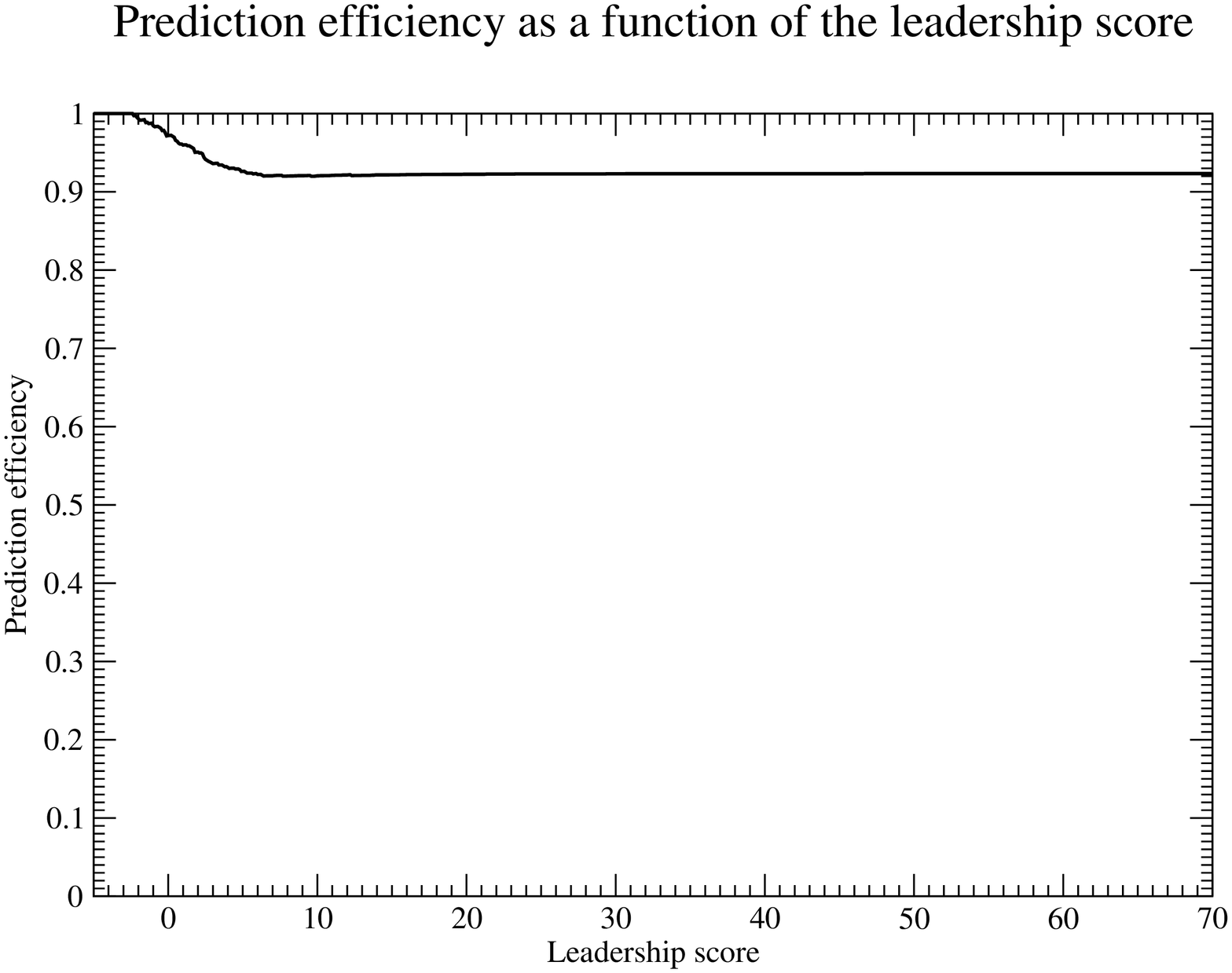}%
\end{psfrags}}%
\caption[Effectiveness of the prediction formula]{Effectiveness of the prediction formula \eqref{predic}. Parameters: $N=900$, $\r=0.85$, $L=7$, $r=0.2$, $\Delta V^{*}=0.1$, $v=5$ $\frac{1}{\textrm{ms}}$, $\tau=100$ ms, $\w=0.1$, $\sigma=0.0081$, $\omega=0.001$ s, $n_\b=100$, $\delta t_\b=15$ ms and $\delta t_{\p}=5$ ms.}\label{FigPredictAll}%
\end{figure}

Now, considering a given neural network (that we did not use to obtain \eqref{predic}) and using
\eqref{predic}, we can predict, even before running our simulation
which neuron will be a leader and which not. Figure
\ref{FigPredict} shows a typical prediction. Remark that this
prediction is pretty reliable because: All the neurons with a
prediction $p_n$ lower than $0$ are not leaders and all the neurons with
a prediction $p_n$ bigger than $6$ are leaders $(\alpha_n>3)$. More precisely, Figure \ref{FigPredictEff} shows the prediction efficiency\footnote{We call prediction efficiency the coefficient $p_e(\alpha)=\frac{\sum_{n|\alpha_n<\alpha}\textrm{correct prediction}_n}{\sum_{n|\alpha_n<\alpha}1}$ where correct prediction$_n$ is $1$ if the neuron $n$ is a leader $\alpha_n>3$ (respectively not a leader $\alpha_n\leq3$) in the simulation and its prediction $p_n>3$ (respectively $p_n\leq3$), otherwise: correct prediction$_n=0$. \label{predeff}} $p_e$ as a function of the leadership score $\alpha$. In Figure \ref{FigPredictEff} we can also remark that the prediction is correct for more than $90\%$ of the neurons $(p_e(\alpha)>0.9 \ \ \forall\alpha)$. This means that Figure \ref{FigPredictEff} shows that the trend of prediction is correct. Of course Figure \ref{FigPredictEff} also shows that there are some mistakes in the predictions especially for prediction $\left(p_n\right)$ between $0$ and $6$ (see Figure \ref{FigPredict}). It is important to remember  that the leaders are function of the dynamical parameters and our coefficients in \eqref{predic} (which are only computed at the linear order) are not.

\begin{figure}[ht]
\begin{center}
\parbox{0.47\textwidth}{%
\begin{psfrags}%
\psfrag{Prediction efficiency as a function of the leadership score}[c][c][1]{\scriptsize Prediction efficiency as a function of the leadership score}%
\psfrag{Prediction efficiency}[c][t][1]{ \tiny Prediction efficiency $p_e(\alpha)$}%
\psfrag{Leadership score}[c][b][1]{ \tiny Leadership score $\left(\alpha\right)$}%
\psfrag{A}[c][c][1]{\tiny A}%
\psfrag{B}[c][c][1]{\tiny {\color{red}B}}%
\psfrag{C}[c][c][1]{\tiny {\color{blue}C}}%
\psfrag{D}[c][c][1]{\tiny {\color{brun}D}}%
\psfrag{E}[c][c][1]{\tiny {\color{green}E}}%
\psfrag{F}[c][c][1]{\tiny {\color{magenta}F}}%
{\includegraphics[width=0.36\textwidth,angle=-90]{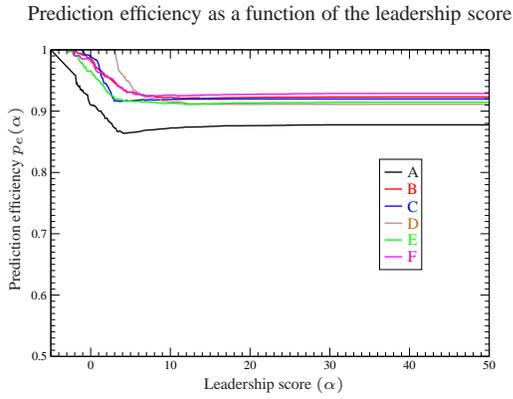}}%
\end{psfrags}
}\parbox{0.53\textwidth}%
{\caption[The prediction efficiency as a function of the leadership score]{ The prediction efficiency $p_e(\alpha)=$ $\frac{\sum_{n|\alpha_n<\alpha}\textrm{correct prediction}_n}{\sum_{n|\alpha_n<\alpha}1}$
as a function of the leadership score $\alpha$ for different simulations (see Figure \ref{FigPredictEff}).
{\bf A) $\mathbf{r=0}$, $\mathbf{\w=0.0666}$, $\mathbf{\Delta V^{*}=0.1}$, $\mathbf{\sigma=0.008}$ and $\mathbf{\omega=0.001}$ s},
{\color{red}B) $r=0.2$, $\w=0.1$, $\Delta V^{*}=0.05$, $\sigma=0.008$ and $\omega=0.001$ s},
{\color{blue}C) $r=0.2$, $\w=0.1$, $\Delta V^{*}=0.1$, $\sigma=0.014$ and $\omega=0.002$ s},
{\color{brun}D) $r=0.2$, $\w=0.1$, $\Delta V^{*}=0$, $\sigma=0.008$ and $\omega=0.001$ s},
{\color{green}E) $r=0.2$, $\w=0.1$, $\Delta V^{*}=0.2$, $\sigma=0.007$ and $\omega=0.001$ s},
{\color{magenta}F) $r=0.2$, $\w=0.1$, $\Delta V^{*}=0.1$, $\sigma=0.008$ and $\omega=0.001$ s},
Common parameters: $N=900$, $\r=0.85$, $L=7$, $v=5$ $\frac{1}{\textrm{ms}}$,
$\tau=100$ ms, $n_\b=100$, $\delta t_\b=15$ ms and
$\delta t_{\p}=5$ ms.}\label{FigPredictGen}}
\end{center}
\end{figure}

Figure \ref{FigPredictGen} shows the prediction efficiency$^{\ref{predeff}}$ as a function of the leadership score $\alpha$. Most of the simulations presented in Figure \ref{FigPredictGen} show that the trend of prediction of \eqref{predic} is correct $\left(p_e(\alpha)\geq0.9,\ \forall\alpha\right)$. Precisely, for a range of simulations with different standard deviation of the membrane potential firing threshold $(\Delta V^{*})$ (Figure \ref{FigPredictGen} simulations {\color{red}B}, {\color{brun}D}, {\color{green}E} and {\color{magenta}F}), for different mean noise $(\sigma)$ and exponential clock rate $(\omega)$ (Figure \ref{FigPredictGen} simulations {\color{blue}C} and {\color{magenta}F})
the trend of prediction is correct. So by looking only at the simulations {\color{red}B}, {\color{blue}C}, {\color{brun}D}, {\color{green}E} and {\color{magenta}F} of Figure \ref{FigPredictGen}, we can say that \eqref{predic} give a pretty good prediction (even without adapting the coefficient of \eqref{predic}
with the dynamical parameters). But for the simulation {\bf A} the trend of prediction is less good $\left(p_e(\alpha)\leq0.9,\ \forall\alpha>0\right)$. The fact is the following, for the simulation {\bf A}, the inhibitory proportion $r=0$. This means that for simulation {\bf A} we did not stay in the parameters range that we gave in Section \ref{parameters} and that we used to find out \eqref{predic}. We can say that the fact that \eqref{predic} does not give a very good trend of predictions for the simulation {\bf A} is a kind of edge effect.
With $\vec x_{r=0}=\left(-0.8927,\ -3.23,\ 0.07,\ \lambda_1, \ -0.02,\ \lambda_2\right)^T$ in \eqref{predic} (where $\lambda_1,\lambda_2\in\real$), we obtain a much better prediction for simulation {\bf A} $\left(p_e(\alpha)\geq0.92,\ \forall\alpha\right)$. Remark that the first number $(-0.8927)$ (the coefficient of $P_{n1}=\textrm{excitatory}_n$) of $\vec x_{r=0}$ has no physical meaning because all neurons are excitatory neurons (if $r=0$ then $P_{n1}=\textrm{excitatory}_n=1$ $\forall n$). For the same reason $\lambda_1,\lambda_2$ in $\vec x_{r=0}$ can be any real number because $P_{n4}$ ($=$\# inhibitory son$_n$) and $P_{n6}$ ($=$\# inhibitory fathers$_n$) are always $0$ if all neurons are excitatory neurons. All this means that the relative importance of the leader properties change by changing the parameters of the simulation but the trend in \eqref{predic} is correct. The sign of each coefficient in \eqref{predic} is robust (as soon as the coefficient has a physical meaning).

We can also explain why we do not use a bigger
matrix $P$. Considering a given neural network, we can extract a
lot of different properties for all neurons $n$: the number of
fathers, the number of sons, the number of fathers'fathers (second
order property\footnote{This means that we need to compute $W^2$
to extract this property.}), the number of loop order
$2,3,4,5\ldots$ in which the neuron $n$ participates,\ldots We can also
reduce the network to an excitatory network and so on. We can as
well try to improve our results by using a lot of tricks and/or
cut off. But using the properties at the first order seem to be an
easy and logical way of doing. Especially considering two facts:
1) A son will not necessarily fire after one of his fathers did, so
what to say about the sons of this son? 2) The leadership score
depends on the dynamical parameters. Taking that fact into account improves more the result than considering the second order properties in the network.

To conclude, \eqref{predic} cannot be taken as a general law and
\eqref{predic} is only a good approximation of the leadership
score in the range of parameters we chose in Section
\ref{parameters}. However, we pretend that the signs in
\eqref{predic} are robust under the change of parameters. This means that our typical leader
profile is the right one.

We pretend that even if \eqref{predic} are computed on small scale simulations, the trend of the typical leader profile is the correct one. Even more we propose to use the following leader profile if the
mean number of connections per neuron vary
\begin{eqnarray}\label{predicAll}
\textrm{if $r,\Delta V^{*}>0$: }\ \ \ p_n&=&
\begin{array}{{l}}
1.22\cdot\textrm{excitatory}_n-2.33\cdot\frac{V_n^*-\langle V^{*}\rangle}{\Delta V^{*}}+1.14\cdot\frac{\textrm{\# excitatory sons}_n}{(1-r)\langle\textrm{\# connections}/\textrm{neuron}\rangle}\\[2mm]-0.42\cdot\frac{\textrm{\# inhibitory sons}_n}{r\ \langle\textrm{\# connections}/\textrm{neuron}\rangle}-1.14\cdot\frac{\textrm{\# excitatory fathers}_n}{(1-r)\langle\textrm{\# connections}/\textrm{neuron}\rangle}\,.
\end{array}
\end{eqnarray}
This equation was computed knowing that in \eqref{predic}: $\langle\textrm{\# connections}/\textrm{neuron}\rangle=11$.

We have already explained in Section \ref{leadersexistence} that
we observe leaders in our simulations as soon as we use reasonable
parameters. Further more,
considering a given simulation, we can store the neurons using the
leadership score. After that, if we decide to remove some of the
(best) leaders from the network and then run again our simulation
we still observe leaders (as good as before). Finally, in a leaky
integrate and fire neuron model there are leaders in the networks
and the neurons which fit better with the typical leader profile
are the leaders.

\subsection{First to fire neurons}\label{firsttofire}

In this section we characterize the first to fire neurons. From this point, we will call first to fire neurons the neurons that fire in a short time after the trigger. This means that the first to fire neurons fire roughly in the successful pre-burst.

\begin{figure}[ht]
\begin{center}
\parbox{0.47\textwidth}{%
\begin{psfrags}%
\psfrag{Number of fathers/sons as a function of the position in the burst}[c][c][1]{\scriptsize \# {\bf fathers}/{\color{blue}sons} as a function of the position in the burst}%
\psfrag{Mean number of fathers/sons per neuron}[c][t][1]{\tiny Mean \# {\bf fathers}/{\color{blue}sons} per neuron}%
\psfrag{Time position in the burst [ms]}[c][b][1]{\tiny Time position in the burst [ms]}%
\psfrag{Fathers}[c][c][1]{\tiny {\bf Fathers}}%
\psfrag{Sons}[c][c][1]{\tiny {\color{blue}Sons}}%
\psfrag{Mean number of connections per}[c][c][1]{\tiny {\color{red}$\langle\textrm{\# connections}/\textrm{neuron}\rangle$}}%
\psfrag{Fa mean fathers activity ratio equation}[c][c][1]{\tiny {\color{brun}$f_a=\sum\textrm{\# fathers}_n\cdot q_n$}}%
{\includegraphics[width=0.36\textwidth, angle=-90]{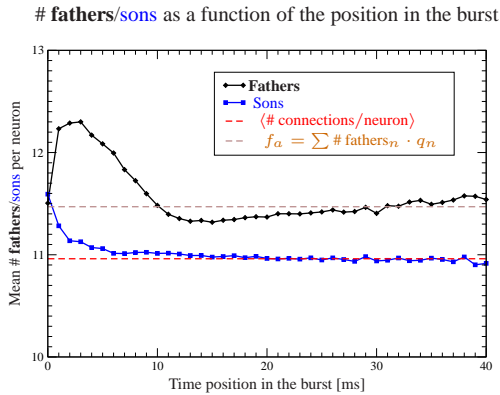}}%
\end{psfrags}
}\parbox{0.53\textwidth}%
{\caption[The mean number of fathers/sons as a function of the position of the spike 1]{The mean number of {\bf fathers}/{\color{blue}sons} per neuron as a function of the time position of a spike of this neuron in the burst. The time bin in which we average the number of {\bf fathers}/{\color{blue}sons} per neuron is $1$ ms. The time position $0$ is reserved for the trigger of the burst. The mean number of connections per neuron in the network (noted {\color{red}$\langle\textrm{\# connections}/\textrm{neuron}\rangle$}) is about $11$. The typical number of fathers of an active neuron {\color{brun}$f_a$} is about $11.5$ (see \eqref{namf} for definition). We obtained this figure by averaging over more than $35000$ bursts. Parameters: $N=900$, $\r=0.85$, $L=7$, $r=0.2$, $\Delta V^{*}=0.05$, $v=5$ $\frac{1}{\textrm{ms}}$, $\w=0.1$, $\tau=100$ ms, $\sigma=0.008$, $\omega=0.001$ s, $n_\b=150$, $\delta t_\b=15$ ms and $\delta t_{\p}=5$ ms.}\label{FigFirstFire}}
\end{center}
\end{figure}

Figure \ref{FigFirstFire} shows the mean number of fathers (or sons) per neuron as a function of the time position of a spike of this neuron in the burst. To obtain Figure \ref{FigFirstFire} we average over all the bursts of a given simulation without taking into account if the trigger of the burst was a leader or not. Nevertheless, when we look at the mean number of sons per trigger\footnote{The time position $0$ in Figure \ref{FigFirstFire} is reserved for the trigger of the burst.} we remark that the trigger of the burst has, on average, more sons than their followers. This fact is consistent with our previous analysis. Remember that Figure \ref{FigLeaderScoreBoss} shows that leaders trigger more bursts than other neurons. This means that the trigger is, on average, a leader. Remember also that \eqref{predic} tells that a leader has typically many excitatory sons (this means a lot of sons as well).

Figure \ref{FigFirstFire} also shows that the first to fire neurons have, on average, more sons than their followers and have, on average, less sons than the trigger. This means that very often several leaders are present in the successful pre-burst (this is linked with the triggering force of leaders, data not shown but see \cite{MaThese}). So the fact that first to fire neurons have typically a lot of sons is consistent with \eqref{predic}.

In Figure \ref{FigFirstFire} the mean number of sons per neuron tends to the mean number of connections per neuron in the network. This means that, except fot the trigger and the first to fire neurons, the mean number of sons per neuron for the neurons which fire during the burst itself is not relevant.

Figure \ref{FigFirstFire} shows explicitly that the mean number of fathers per neuron of the first to fire neurons is higher than the mean number of connections per neuron in the network. This means that the first to fire neurons are characterized by a high number of fathers. This is consistent with the fact that neurons with a lot of fathers light up early in the burst, respectively in the successful pre-burst (see \cite{TsviQuorum}). In facts neurons with a lot fathers have an higher neural activity than neurons with a few fathers (data not shown). And again, this is consistent with \cite{TsviQuorum} and Figure \ref{FigFirstFire} which shows that active neurons in the burst are, on average, neurons with a lot of fathers (typically more fathers than the mean number of connections per neuron in the network).

In Figure \ref{FigFirstFire} the mean number of fathers per neuron (unlike the mean number of sons) does not tend to the mean number of connections per neuron in the network but seems to increase again after the decay of the first to fire neurons. This is consistent with the following fact: in a burst, after a while, neurons are able to fire again (as soon as their membrane potentials are charged enough). And, of course, the neurons with a lot of fathers are able to fire again sooner than the others.

What about the mean number of fathers of the trigger? Figure \ref{FigFirstFire} shows a tendency of the trigger to have more fathers than the mean number of connections per neurons. This last fact seems to contradict our predictions \eqref{predic}. But in Figure \ref{FigFirstFire} the mean number of fathers of the trigger is approximately what we call the typical number of fathers of an active neuron $f_a$:
\begin{equation}\label{namf}
 f_a=\sum_{n=1}^{N}\textrm{\# fathers}_n\cdot q_n
\end{equation}
where $\textrm{\#}$ fathers$_n$ is the number of fathers of neuron $n$ and $q_n$ is the neural activity ratio of neuron $n$. Figure \ref{FigFirstFire} shows that (except for the first to fire) the mean number of fathers per neuron is about $f_a$ the typical number of fathers of an active neuron and shows that the number of fathers of the typical trigger is not different from $f_a$. These two facts can explain the term concerning the number of fathers in \eqref{predic}, but still, we can think it contradicts \eqref{predic}.

\begin{figure}[ht]
\begin{center}
\parbox{0.47\textwidth}{%
\begin{psfrags}%
\psfrag{Number of fathers/sons as a function of the position in the burst}[c][c][1]{\scriptsize \# {\bf fathers}/{\color{blue}sons} as a function of the position in the burst}%
\psfrag{Mean number of fathers/sons per neuron}[c][t][1]{\tiny Mean \# {\bf fathers}/{\color{blue}sons} per neuron}%
\psfrag{Fathers}[c][c][1]{\tiny {\bf Fathers}}%
\psfrag{Sons}[c][c][1]{\tiny {\color{blue}Sons}}%
\psfrag{Mean number of connections per}[c][c][1]{\tiny {\color{red}$\langle\textrm{\# connections}/\textrm{neuron}\rangle$}}%
\psfrag{Fa mean fathers activity ratio equation}[c][c][1]{\tiny {\color{brun}$f_a=\sum\textrm{\# fathers}_n\cdot q_n$}}%
\psfrag{Time position in the burst [ms]}[c][b][1]{\tiny Time position in the burst [ms]}%
{\includegraphics[width=0.36\textwidth,angle=-90]{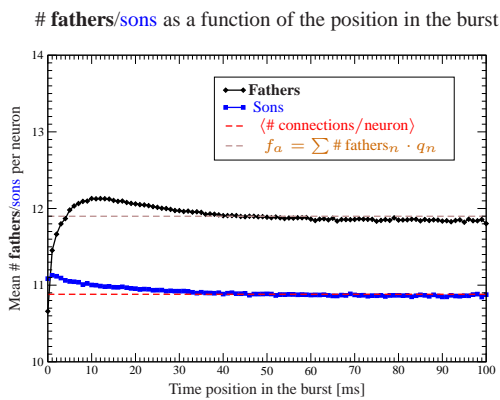}}%
\end{psfrags}
}\parbox{0.53\textwidth}%
{\caption[The mean number of fathers/sons as a function of the position of the spike 2]{The mean number of {\bf fathers}/{\color{blue}sons} per neuron as a function of the time position of a spike of this neuron in the burst. The time bin in which we average the number of {\bf fathers}/{\color{blue}sons} per neuron is $1$ ms. The time position $0$ is reserved for the trigger of the burst. The mean number of connections per neuron in the network (noted {\color{red}$\langle\textrm{\# connections}/\textrm{neuron}\rangle$}) is about $11$. The typical number of fathers of an active neuron {\color{brun}$f_a$} is about $11.9$ (see \eqref{namf} for definition). We obtained this figure by averaging over more than $35000$ bursts.
Parameters: $N=900$, $\r=0.85$, $L=7$, $r=0.2$, $\Delta V^{*}=0.2$, $v=5$ $\frac{1}{\textrm{ms}}$, $\w=0.1$, $\tau=100$ ms, $\sigma=0.007$, $\omega=0.001$ s, $n_\b=150$, $\delta t_\b=15$ ms and $\delta t_{\p}=5$ ms.}\label{FigEffectFather}}
\end{center}
\end{figure}

To clarify this conflict, let us have a look at Figure \ref{FigEffectFather}. This figure shows, as Figure \ref{FigFirstFire}, the mean number of fathers (and sons) per neuron as a function of the time position of a spike of this neuron in the burst. In Figure \ref{FigEffectFather} the standard deviation of the membrane potential firing threshold $\Delta V^{*}$ is much higher than the one in Figure \ref{FigFirstFire}. In Figure \ref{FigEffectFather}, we see that the number of sons of the trigger as well as the one of the first to fire neurons is higher than the mean number of connections per neuron in the network. And we see that the number of fathers of the trigger is lower than the mean number of connections per neuron in the network. But the number of fathers of the first to fire neurons is higher than the mean number of connections per neuron in the network. This force us to conclude that the simulation shown in Figure \ref{FigEffectFather} does not contradict \eqref{predic}. The reason for that is that the membrane potential firing threshold plays a more important r{\^o}le in this case.

We can also remark that in Figure \ref{FigFirstFire} like in Figure \ref{FigEffectFather}, the mean number of fathers per neuron tends approximately to the typical number of fathers of an active neuron $f_a$ while we look forward in time in the burst.

Now, let us clarify the observation of Figures \ref{FigLeaderScore} concerning the neurons 634 and 643. Neuron 643 is one of the best leaders of the simulation of Figure \ref{FigLeaderScoreAll} while restricting the observation to $60$ neurons only implies that neuron 634, which is not a leader (see Figure \ref{FigLeaderScoreAll}), appears as a leader (see Figure \ref{FigLeaderScore60}). The fact is that both neurons 634 and 643 have the profile of first to fire neurons (mean number of fathers close to the typical number of fathers of an active neuron). But the membrane potential firing threshold of the two neurons explain why only neuron 643 can be a leader: $V^{*}_{\textrm{634}}=1.009$ and $V^{*}_{\textrm{643}}=0.869$ $\left(\langle V^{*}\rangle=1\textrm{ and }\Delta V=0.05\right)$. So while restricting the observation to $60$ neurons only, it is possible that the first to fire neurons appear as leaders. In the experiments \cite{Leadership} it was impossible to distinguish between true leaders and first to fire neurons.

Finally we have a competition between two facts: To be a leader, a neuron needs to be able to fire early (in order to do that it must have a low membrane potential firing threshold and a lot of fathers) and to be a leader, a trigger must trigger more bursts than we expect. Typically a leader has a good triggering power (this means a lot of excitatory sons) and a relatively low neural activity ratio (this means less fathers than the typical number of fathers of an active neuron). That is the message of \eqref{predic} (respectively \eqref{predicAll}). Remember that \eqref{predic} was obtained by averaging over simulations, so \eqref{predic} express only a tendency of what kind of neurons the leaders are.

\section{Conclusion}

Experimental studies show the existence of leader neurons in
population bursts of 2D living neural networks
\cite{LeaderMarom,Leadership}. This leads naturally to a question:
Do leaders also exist in neural network simulations? In this study, we have
proved that stable leaders exist in leaky integrate and
fire neural network simulations.

In our simulations, we saw that leaders depend on the network but
also on the dynamical parameters. Because the leaders mainly
depend on the network and on the own properties of the neurons, we
were able to find some important properties for these leaders.
Firstly, they are excitatory neurons and have a low membrane potential
firing threshold (ability of firing early). Secondly, they send a signal to a lot of
excitatory neurons and to a few inhibitory neurons (triggering power). Finally, they
receive only a few signal from other excitatory neurons (trigger more bursts than we expect by looking at the neural activity).

The study of more biological plausible models \cite{IzhiWhich,WulfCompare}
(like Hodgkin-Huxley) would be very interesting too and could
probably also produce stable leaders. These models could probably be used
to find out other leader properties.

\section*{Acknowledgments}\addcontentsline{toc}{section}{Acknowledgments}

I would like to thank Jean-Pierre Eckmann, my PhD advisor at the
University of Geneva, Switzerland, for his continuous useful help along this study. This paper could not have been without the help of Elisha Moses, Weizmann Institute of Science,
Israel. During all this research, I was supported by the Fonds National
Suisse. Finally I would like to thank Sonia Iva Zbinden.

\addcontentsline{toc}{section}{Bibliography}
\bibliographystyle{unsrt}
\bibliography{lIF_Zbinden}

\begin{thebibliography}{10}

\bibitem{LeaderMarom}
D.~Eytan and S.~Marom.
\newblock {Dynamics and effective topology underlying synchronization in
  networks of cortical neurons}.
\newblock {\em J. Neurosci.}, 26(33):8465--8476, 2006.

\bibitem{Leadership}
J.-P. Eckmann, S.~Jacobi, S.~Marom, E.~Moses, and C.~Zbinden.
\newblock {Leader neurons in population bursts of 2D living neural networks}.
\newblock {\em New J. Phys.}, 10(015011), 2008.

\bibitem{GerstnerKistler}
W.~Gerstner and W.~M. Kistler.
\newblock {\em {Spiking neuron models. Single neurons, populations,
  plasticity}}.
\newblock Cambridge University press, 2002.

\bibitem{Cessac}
B.~Cessac.
\newblock {A discrete time neural network model with spiking neurons. Rigorous
  results on the spontaneous dynamics}.
\newblock {\em Journal of Mathematical Biology}, 56(3):311--345, 2008.

\bibitem{IzhiWhich}
E.~M. Izhikevich.
\newblock {Which model to use for cortical spinking neurons?}
\newblock {\em IEEE transactions on neural networks}, 15(5):1063--1070, 2004.

\bibitem{WulfCompare}
R.~Naud and W.~Gerstner.
\newblock {How good are neuron models?}
\newblock {\em Science}, 326:379--380, 2009.

\bibitem{TscherterInitiation}
A.~Tscherter, M.O. Heuschkel, P.~Renaud, and J.~Streit.
\newblock {Spatiotemporal characterization of rhythmic activity in rat spinal
  cord slice cultures}.
\newblock {\em Eur J Neurosci}, 14(2):179--190, 2001.

\bibitem{MaedaGeneration1995}
E.~Maeda, H.P. Robinson, and A.~Kawana.
\newblock {The mechanisms of generation and propagation of synchronized
  bursting in developing networks of cortical neurons}.
\newblock {\em J Neurosci}, 15(10):6834--6845, 1995.

\bibitem{droge1986mac}
M.H. Droge, G.W. Gross, M.H. Hightower, and L.E. Czisny.
\newblock {Multielectrode analysis of coordinated, multisite, rhythmic bursting
  in cultured CNS monolayer networks}.
\newblock {\em J Neurosci}, 6(6):1583--1592, 1986.

\bibitem{PotterDevelopment}
D.A. Wagenaar, J.~Pine, and S.M. Potter.
\newblock {An extremely rich repertoire of bursting patterns during the
  development of cortical cultures}.
\newblock {\em BMC Neurosci}, 7(11), 2006.

\bibitem{Wagenaar2006}
D.A. Wagenaar, J.~Pine, and S.M. Potter.
\newblock {Searching for plasticity in dissociated cortical cultures on
  multi-electrode arrays}.
\newblock {\em J Neg Res BioMed}, 5(1):16, 2006.

\bibitem{JordiDevelopment}
J.~Soriano, M.~Rodriguez Martinez, T.~Tlusty, and E.~Moses.
\newblock {Development of input connections in neural cultures}.
\newblock {\em Proc Natl Acad Sci USA}, 105(37):13758--13763, 2008.

\bibitem{Bootstrap}
J.-P. Eckmann and T.~Tlusty.
\newblock {Remarks on bootstrap percolation in metric networks}.
\newblock {\em J. Phys. A: Math. Theor}, 42(205004), 2009.

\bibitem{CessacNumerical}
B.~Cessac, O.~Rochel, and T.~Viéville.
\newblock {Introducing numerical bounds to event-based neural network
  simulation}.
\newblock {\em arXiv}, (0810.3992), 2008.

\bibitem{MaThese}
C.~Zbinden.
\newblock {\em Leader Neurons in Living Neural Networks and in Leaky Integrate
  and Fire Neuron Models}.
\newblock PhD thesis, University of Geneva, 2010.
\newblock \url{http://archive-ouverte.unige.ch/unige:5451}.

\bibitem{CessacSam}
B.~Cessac and M.~Samuelides.
\newblock {From Neuron to Neural Network Dynamics}.
\newblock {\em EPJ Special Topics}, 142(1):7--88, 2007.

\bibitem{RevueNeurones}
L.~Gruendlinger, J.-P. Eckmann, O.~Feinerman, J.~Soriano, and E.~Moses.
\newblock {The Physics of Living Neural Networks}.
\newblock {\em Physics reports}, 449:54--76, 2007.

\bibitem{MosesAlvarez}
E.~Alvarez-Lacalle and E.~Moses.
\newblock {Slow and fast waves in 1-D cultures of excitatory neurons}.
\newblock 2007.

\bibitem{Email}
J.-P. Eckmann, E.~Moses, and D.~Sergi.
\newblock {Entropy of dialogues creates coherent structures in e-Mail traffic}.
\newblock {\em PNAS}, 101(40):14333--14337, 2004.

\bibitem{MultiInf}
N.~Ay and A.~Knauf.
\newblock {Maximizing multi-information}.
\newblock {\em Kybernetika}, 42(5):517--538, 2006.

\bibitem{WulfAdaptExpo}
R.~Naud, N.~Marcille, C.~Clopath, and W.~Gerstner.
\newblock Firing patterns in the adaptive exponential integrate-and-fire model.
\newblock {\em Biological Cybernetics}, 99:335--347, 2008.

\bibitem{WulfAdaptMain}
R.~Brette and W.~Gerstner.
\newblock {Adaptive Exponential Integrate-and-Fire Model as an Effective
  Description of Neuronal Activity}.
\newblock {\em J Neurophysiol}, 94:3637--3642, 2005.

\bibitem{WulfPredict}
H.-R.~Luscher R.~Jolivet, A.~Rauch and W.~Gerstner.
\newblock {Predicting spike timing of neocortical pyramidal neurons by simple
  threshold models}.
\newblock {\em Journal of Computational Neuroscience}, 21:35--49, 2006.

\bibitem{LinAlg}
J.H. Wilkinson and C.~Reinsch.
\newblock {\em {Linear algebra}}.
\newblock Springer, 1971.

\bibitem{TsviQuorum}
O.~Cohen, A.~Keselman, E.~Moses, M.~Rodriguez Martinez, J.~Soriano, and
  T.~Tlusty.
\newblock {Quorum percolation: More is different in living neural networks}.
\newblock {\em submitted}, 2009.

\end{thebibliography}

\end{document}